\documentclass[
 reprint,
superscriptaddress,
 amsmath,amssymb,
 aps,
]{revtex4-2}

\usepackage{graphicx}
\usepackage{float}

\usepackage{dcolumn}
\usepackage{bm}
\usepackage{verbatim}
\usepackage{xcolor}
\usepackage{subfig}
\usepackage{lipsum}
\usepackage{hyperref}
\usepackage{cleveref}
\crefname{figure}{Fig.}{Figs.}

\usepackage{siunitx}
\usepackage[inkscapeformat=png]{svg}

\begin{document}
\title{SmartTrap: Automated Precision Experiments with Optical Tweezers}

\author{Martin Selin}
\affiliation{Physics Department, University of Gothenburg, 412 96 Gothenburg, Sweden}
\author{Antonio Ciarlo}
\affiliation{Physics Department, University of Gothenburg, 412 96 Gothenburg, Sweden}

\author{Giuseppe Pesce}
\affiliation{Physics Department, University of Gothenburg, 412 96 Gothenburg, Sweden}
\affiliation{Department of Neuroscience, Reproductive Sciences and Dentistry, Universit\`a degli studi di Napoli “Federico II”  Naples, 80131, Italy}
\affiliation{Department of Physics, Universit\`a degli studi di Napoli “Federico II” Complesso Universitario Monte S. Angelo, Naples, 80126 Napoli, Italy}
\author{Lars Bengtsson}
\affiliation{Physics Department, University of Gothenburg, 412 96 Gothenburg, Sweden}

\author{Joan Camunas-Soler}
\affiliation{Department of Medical Biochemistry and Cell Biology, Institute of Biomedicine, University of Gothenburg, Gothenburg 405 30, Sweden}
\affiliation{Wallenberg Centre for Molecular and Translational Medicine, Sahlgrenska Academy, University of Gothenburg, 405 30 Gothenburg, Sweden}

\author{Vinoth Sundar Rajan}
\affiliation{Department of Life Sciences, Chalmers University of Technology, 41296 Gothenburg, Sweden}
\affiliation{Department of Chemistry and Chemical Engineering, Chalmers University of Technology, 41296 Gothenburg, Sweden}

\author{Fredrik Westerlund}
\affiliation{Department of Life Sciences, Chalmers University of Technology, 41296 Gothenburg, Sweden}

\author{L. Marcus Wilhelmsson}
\affiliation{Department of Chemistry and Chemical Engineering, Chalmers University of Technology, 41296 Gothenburg, Sweden}

\author{Isabel Pastor}
\affiliation{Small Biosystems Lab, Condensed Matter Physics Department, University of Barcelona, 08028 Barcelona, Spain}

\author{Felix Ritort}
\affiliation{Small Biosystems Lab, Condensed Matter Physics Department, University of Barcelona, 08028 Barcelona, Spain}
\affiliation{Institut de Nanociència i Nanotecnologia (IN2UB), Universitat de Barcelona, Spain}
\affiliation{Reial Acadèmia de Ciències i Arts de Barcelona (RACAB), La Rambla 115, 08002 Barcelona, Spain}

\author{Steven B. Smith}
\affiliation{Steven B. Smith Engineering, Los Lunas, New Mexico, USA}

\author{Carlos Bustamante}
\affiliation{Department of Molecular and Cell Biology, University of California, Berkeley, Berkeley, CA 94720, USA}
\affiliation{California Institute for Quantitative Biosciences, University of California, Berkeley, Berkeley, CA 94720, USA}
\affiliation{Jason L. Choy Laboratory of Single-Molecule Biophysics, University of California, Berkeley, Berkeley, CA, USA}
\affiliation{Department of Physics, University of California, Berkeley, Berkeley, CA, USA}
\affiliation{Howard Hughes Medical Institute, University of California, Berkeley, Berkeley, CA, USA}
\affiliation{Kavli Energy Nanoscience Institute, University of California, Berkeley, Berkeley, CA, USA}

\author{Giovanni Volpe}
\affiliation{Physics Department, University of Gothenburg, 412 96 Gothenburg, Sweden}
\affiliation{SciLifeLab, Physics Department, University of Gothenburg, 412 96 Gothenburg, Sweden}

\date{May 8, 2025}
\begin{abstract}

There is a trend in research towards more automation using smart systems powered by artificial intelligence. While experiments are often challenging to automate, they can greatly benefit from automation by reducing labor and increasing reproducibility.
For example, optical tweezers are widely employed in single-molecule biophysics, cell biomechanics, and soft matter physics, but they still require a human operator, resulting in low throughput and limited repeatability.
Here, we present a smart optical tweezers platform, which we name SmartTrap, capable of performing complex experiments completely autonomously. 
SmartTrap integrates real-time 3D particle tracking using deep learning, custom electronics for precise feedback control, and a microfluidic setup for particle handling. 
We demonstrate the ability of SmartTrap to operate continuously, acquiring high-precision data over extended periods of time, through a series of experiments.
By bridging the gap between manual experimentation and autonomous operation, SmartTrap establishes a robust and open source framework for the next generation of optical tweezers research, capable of performing large-scale studies in single-molecule biophysics, cell mechanics, and colloidal science with reduced experimental overhead and operator bias.
\end{abstract}
\maketitle

\newpage

\section*{Introduction}

Artificial Intelligence (AI) has advanced at a staggering pace, giving us from conversational chatbots to foundation models such as AlphaFold that have revolutionized automated protein structure prediction and design~\cite{lu2024ai,jumper2021highly,DLCC}.
Although such breakthroughs have benefited many computational domains, the integration of AI into experimental sciences still faces significant hurdles, not least due to variability between experimental protocols~\cite{holland2020automation}.

Optical tweezers started with Ashkin’s demonstration of particle trapping by radiation pressure in 1970~\cite{ashkin1970acceleration} and of a single beam optical trap in 1986~\cite{ashkin1986observation}. Today, they are widely used in a broad range of fields from physics to
biology and chemistry to exert and measure microscopic forces~\cite{volpe2023roadmap}. 
For example, in physics, they have been employed for trapping and manipulating microscopic particles, enabling precise studies of soft matter and non-equilibrium dynamics~\cite{magazzu2023investigation,turlier2016equilibrium}.
In biophysics, they have enabled precision measurements of the forces involved in the stretching of single molecules~\cite{bustamante2021optical}, of the forces generated by biomolecular motors~\cite{liu2014mechanical}, and of the mechanical properties of membranes~\cite{zhang2008optical}.
In colloidal chemistry, optical tweezers have proven invaluable for probing interparticle interactions~\cite{grier1997optical}, critical Casimir forces~\cite{hertlein2008direct}, and depletion interactions~\cite{crocker1999entropic}.

However, optical tweezers' focus on manipulating single objects also represents a significant limitation, as it results in an inherently low throughput. Experiments are conducted on one particle, molecule, or cell at a time, typically requiring continuous supervision by a trained practitioner. Furthermore, this dependence on manual operation significantly increases the time and cost of collecting large datasets, while also introducing the potential for human bias, reducing reproducibility.

This is starting to change, in large part thanks to the rise of deep learning~\cite{cichos2020machine}, which has recently been used to enhance optical tweezers~\cite{ciarlo2024deep}. 
Various aspects of optical tweezers' experimental procedures have been automated. For example, automated optical tweezers have been used to position particles with high precision to construct crystal-like structures~\cite{melzer2021assembly}, and a combination of real-time image analysis and machine learning has been used to automatically trap and classify particles~\cite{teixeira2023autonomous}.
However, to date, precision measurements, such as experiments with single molecules, single cells, and individual colloidal interactions, have been out of reach for autonomous procedures because of their complexity. Still, it would be very valuable to increase the throughput, especially for complex experiments such as single-molecule experiments, where data is challenging to gather ~\cite{bustamante2021optical}. Automation would also be ideal for studying rare or transient events, such as misfolding of proteins, and for sampling of heterogeneous distributions.

Here, we present a smart optical tweezers platform, which we name SmartTrap, designed to perform advanced experiments without human intervention. This is achieved through a completely digital control over experimental procedures and smart event-driven algorithms. Digital control is made possible by integrating custom-built electronics, microfluidics, and optical controls into a single software system.  Combining this with real-time deep-learning-based image analysis and closed-loop feedback algorithms enables SmartTrap to autonomously react to events and perform complex experimental procedures autonomously. We demonstrate the versatility of SmartTrap across four paradigmatic experiments: particle size characterization, single-molecule DNA stretching to observe force-induced overstretching transitions, optical deformation of red blood cells to probe membrane stiffness, and measurement of electrostatic forces between colloidal particles to characterize short-range interactions. By making both software and hardware open-source, we aim to inspire others to use and build on SmartTrap, thus establishing a framework for smart optical tweezers, transforming the scope of applications in biophysics, cell mechanics, and colloidal science.

\section*{Results}

\subsection*{The optical tweezers system}

The SmartTrap optical tweezers system utilizes two counter-propagating lasers to create an optical trap, with forces measured directly from the momentum change of the trapping beams \cite{smith20037}. We have developed a custom electronics controller for this system, which controls the positioning of the lasers and sample stage while simultaneously measuring the position and deflection of the lasers. SmartTrap also integrates microfluidic pumps, laser power controls, and a digital camera into a single control interface. A more detailed description of the instrument can be found in the ``Optical Tweezers System'' section of the Supplementary Materials, while the key features are outlined below.

\begin{figure*}
    \centering
    \includegraphics[width=0.95\textwidth]{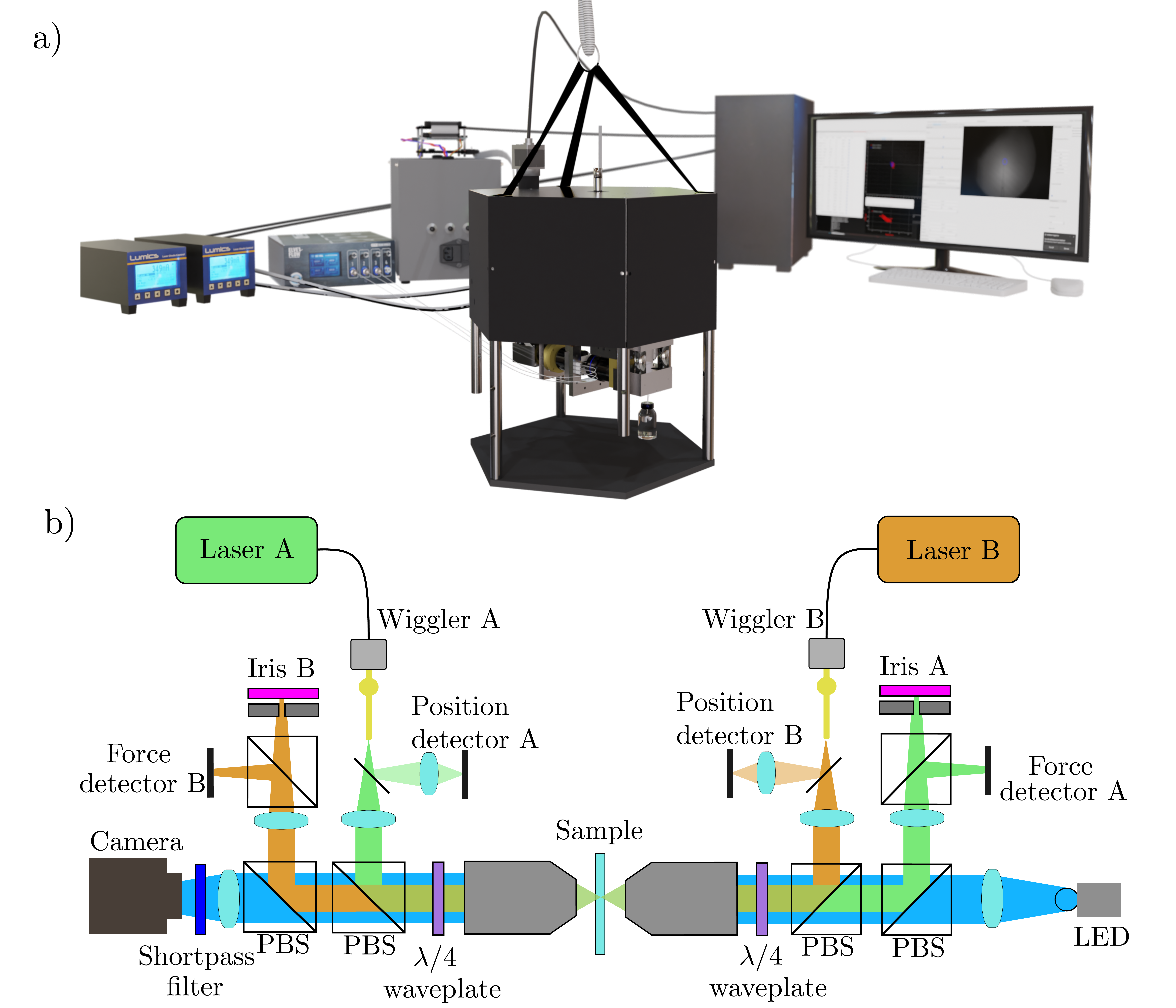}
    \caption{{\bf SmartTrap setup.}
    (\textbf{a}) 3D illustration of the SmartTrap optical tweezers instrument and control system. The instrument is in the front and the control system behind it with the computer and interface on the right and the controllers on the left. 
    (\textbf{b}) Schematics of the optics. Lasers A and B follow mirrored pathways and form a counter-propagating optical trap in the sample. Along each laser path, there are two 2D position sensitive detectors (PSDs): the first monitors the laser position and the second measures the force from the laser on trapped objects from the scattered light. PBS is short for polarizing beam splitter and LED for light emitting diode.}
    \label{fig:miniTweezersSchematics}
\end{figure*}

The SmartTrap setup is illustrated in \textbf{\cref{fig:miniTweezersSchematics}}. The counter-propagating arrangement can be best understood by following the path of laser A. The laser exits the optical fiber through a wiggler. This is a custom actuator designed for 2D positioning of the laser by tilting (wiggling) the optical fiber using piezoelectric actuators, thus moving the lasers in the plane of the sample; see the ``Laser wigglers'' section in the Supplementary Materials. A position sensitive detector (PSD) placed before the sample (``Position detector A'') detects the position of the laser providing a signal that can be exploited in a feedback loop.
After the sample, a second PSD (``Force detector A'') measures the deflection of the laser caused by the optically trapped object allowing the measurement of the lateral momentum transfer and, therefore, the lateral optical force. 
A photodiode with an iris (``Iris A'') measures the size of the laser spot, which is used to determine the force acting on the particle along the axial direction. 
Laser B follows a path that mirrors that of laser A. 
Together, lasers A and B form a single trap where the scattering forces from the two beams cancel out creating a counter-propagating optical trap that is more stable than what a single, highly-focused beam trap would achieve at the same power.

The system makes use of a custom microfluidics chamber with three parallel channels, as illustrated in \textbf{Suppl. \cref{fig:MicrofluidicsChamber}}\textbf{a}. The experiments are performed in the central channel, while the particles to be used in the experiments flow in the other two channels. These side channels are connected to the main channel through capillaries.
The microfluidic chamber contains also a micropipette, connected to a digitally controlled pump, with the tip of the pipette positioned in the middle of the central channel. During experiments with two particles, one of the particles is held by the micropipette using suction as illustrated in \textbf{Suppl. \cref{fig:MicrofluidicsChamber}}\textbf{b}.

The custom electronic controller steers the sample stage and the positions of the lasers, while also reading the various photosensors. This enables us to implement specialized feedback algorithms to control the system. There is very low latency when the algorithms are running on the controller (ca \SI{0.1}{\milli\second}), which results in faster feedback than when sending commands from the computer (ca \SI{10}{\milli\second}). This controller communicates with the host computer using a serial USB protocol connected with a standard USB cable. The laser power supplies and the microfluidics pump are controlled by separate commercial controllers.

All the controllers, the photodiodes, and the camera are connected to a host computer running a custom Python program providing a graphical user interface (GUI). This provides the user with full digital monitoring and control of the SmartTrap and enables synchronized control of its different components. Further details on the software and electronics are outlined in the ``Optical Tweezers System'' section of the Supplementary Materials.

\subsection*{Real time image analysis}

In order to automate the experimental procedures, it is essential to analyze the video feed from the camera. We implemented this analysis using deep learning because of its precision, versatility, and speed~\cite{midtvedt2021quantitative}.

The first step of the image analysis consists of detecting the presence of the particles and the pipette. This is done using an artificial neural network that utilizes the YOLO object detection framework~\cite{redmon2016you} and, in particular, YOLO V5~\cite{yolov5}.
We trained the network using a dataset consisting primarily of synthetic images simulated using the DeepTrack2 software package~\cite{midtvedt2021quantitative,deeptrack2}, for which the ground truth is known exactly.
Furthermore, we also included 1,000 manually annotated experimental images because this improved the detection accuracy, which we attribute to challenges in accurately simulating a micropipette. Examples images from the training dataset, inlcuding the targets, are shown in \textbf{Suppl. \cref{fig:YOLO_training}}.
The network predictions consist of bounding boxes around the objects of interest. The sizes of the boxes give an estimate of the particle sizes and the extent of the micropipette. The box centers provide the positions of the particles in the xy-directions.

For many experiments, the optically trapped particle and the particle held in the pipette need to be in the same plane along the z-direction (axial direction).
This means that also the axial position of the optically trapped particle needs to be accurately estimated. We did this using a convolutional neural network trained with images of the particles simulated with the DeepTrack2 package \cite{midtvedt2021quantitative,deeptrack2}, see \textbf{Suppl. \cref{fig:Z_trainingData}}. 
The convolutional network performs its predictions taking as input a $128\times128$ pixel image of the particle centered on the bounding box determined using the YOLO algorithm.
Predictions on real images are consistent within a range of a few microns. Example predictions for particles in the size range used for the experiments (from 2 to 4~microns in diameter) are shown in \textbf{Suppl. \cref{fig:Z-focuse_real}} and the performance of the networks is shown in Supplementary video 1.

Further details on the neural networks used and how they are trained are described in the ``Artificial Neural Networks'' section of the Supplementary Materials.

Note that the z-prediction also serves a second purpose in determining whether a second particle has entered the trap. If a second particle enters the trap, this will displace the trapped particle along the z-axis in the trap leading to a shift in the z-predictions, which is used to determine if the system has accidentally trapped more than one particle.

\begin{figure*}
    \centering
    \includegraphics[width=0.95\textwidth]{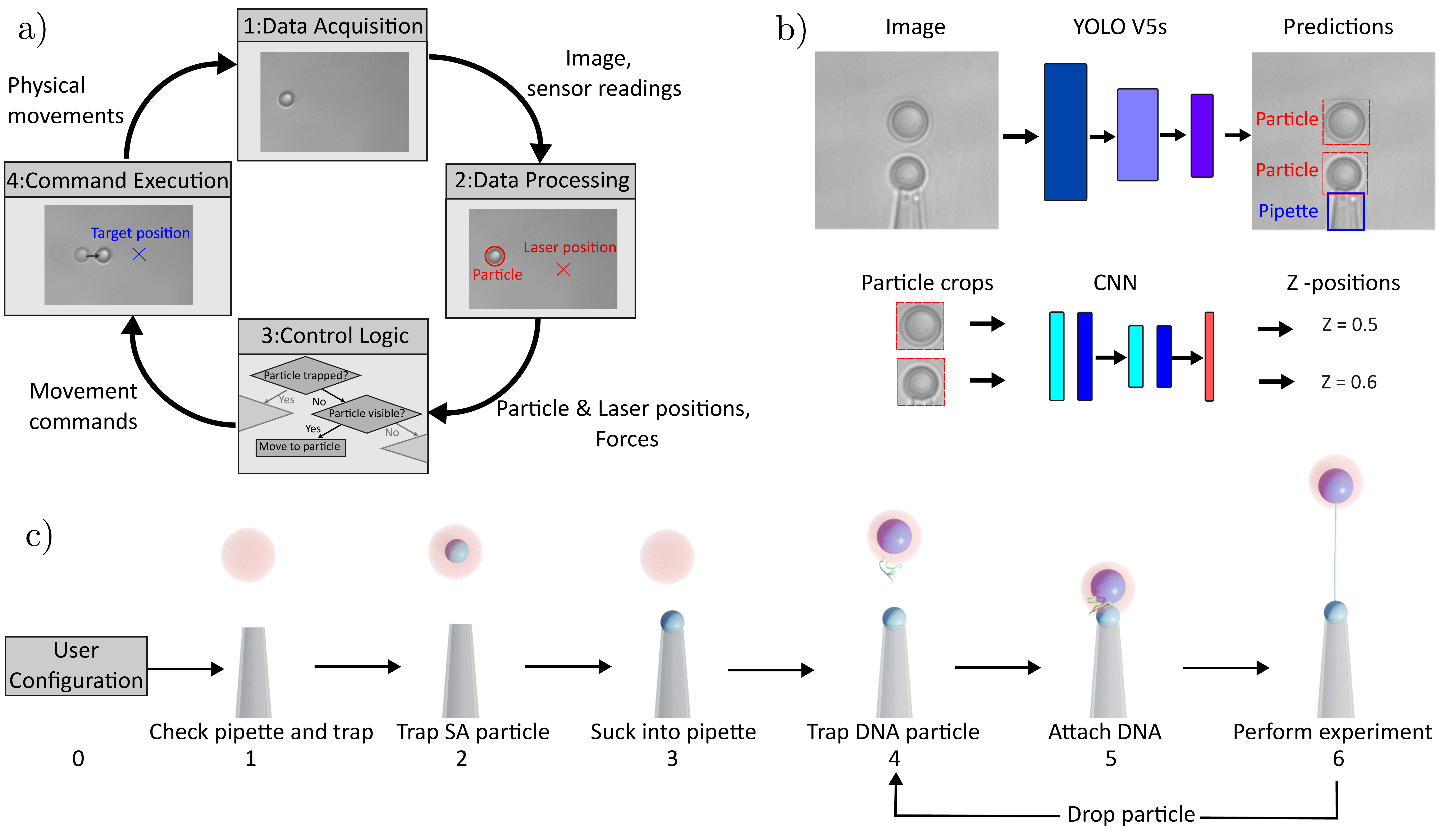}
    \caption{
    {\bf Algorithms used for automation.} 
    (\textbf{a}) Data processing feedback loop. After the data has been acquired (Step 1, Data Acquisition), it is processed to extract information about the experiment (Step 2, Data Processing). This information is then used to make a decision about what to do next (Step 3, Control Logic), which is then executed by the instrument (Step 4, Command Execution).
    (\textbf{b}) Neural networks employed in the image analysis. First, the objects of interest are located with YOLO. Then, the z-positions of the particles are determined with the help of a convolutional neural network (CNN).
    (\textbf{c}) Main steps of an autonomous DNA pulling experiment executed with SmartTrap. It starts with a configuration procedure (step 0),  followed by the system investigating if there is a particle in the pipette and the optical trap (step 1). Step 2 is getting the streptadavin (SA) particle followed by step 3 when the particle is held by suction into the pipette. Then, in step 4 also the particle with attached DNA is collected. In step 5, the system will try to attach the DNA by gently pushing the particles together. Once a DNA molecule is detected, the system performs the pulling experiment (step 6). When the experiment is finished the system will replace the DNA particle by flushing the chamber with buffer solution and returning to step 4.}
    \label{fig:AutomationLoop}
\end{figure*}

\subsection*{Feedback algorithms}

To achieve full autonomous operation, the SmartTrap is controlled by custom algorithms that use the readings from the various sensors for real-time feedback and to keep track of what stage of the experiment is being executed.
This forms a closed-loop system consisting of 4 primary steps, as illustrated in \textbf{\cref{fig:AutomationLoop}a}:
\begin{enumerate}
    \item  \textbf{Data Acquisition:} Data are acquired by reading the various sensors (PSDs, photodiodes, and camera) and the positions of the motors.
    
    \item \textbf{Data Processing:} The latest data are processed to extract the necessary information, such as the location of the particles, the locations of the lasers, and whether a force is acting on the optically trapped particle. Image analysis is performed using the neural networks as illustrated in \textbf{\cref{fig:AutomationLoop}b}.
     
    \item \textbf{Control Logic:} The processed information is used by the instrument to detect events (e.g., a particle entering the trap), enabling it to decide what to do next based on which part of the experimental procedure is being performed, as illustrated for the case of a DNA pulling experiment in \textbf{\cref{fig:AutomationLoop}c}. The decision process is customized for different experiments but many building blocks remain the same (e.g., trapping of particles, alignment). 
    
    \item \textbf{Command Execution:} Commands issued by the control logic are executed by the system, for instance to move the sample stage to a target positions to trap a particle.
\end{enumerate}
Importantly, the data acquisition and the command execution are performed asynchronously with the control logic by a combination of different threads and processes in Python. 
This makes it possible to sample and record data from the various sensor and camera at higher frequencies than the networks can analyze.

When run on a computer with a Nvidia RTX 3090 graphics card, the decision process, including the image analysis, typically operates at about \SI{20}{\hertz}, which is sufficient for real-time feedback and comparable to the camera (Basler a2A5320-23umBAS) frame rate of ca $23$ frames per second when recording at full frame. The vast majority of the computational time in each decision is spent on analyzing the images. Therefore, the feedback rate will depend on the field of view and how many particles are visible, since more particles in view means that the z-network needs to perform more predictions.
We also found that a significant proportion of the analysis is due to latency in sending images and results to and from the GPU by observing only a $50\%$ increase in processing time for the YOLO network when two images in a batch are processed compared to one. Still, the process is more than sufficiently fast to autonomously perform experiments and significantly faster than the typical human reaction time \cite{woods2015factors}.

\subsection*{Particle characterization}

Optical tweezers are well suited for precision sorting because they can manipulate diverse samples in a non-contact manner and provide a variety of input signals about the trapped object. Here, we present an event-driven approach that rapidly characterizes particles using real-time image analysis and force measurements. We use a mixture of two particles, where we identify the particles of one type and measure their hydrodynamic radius.

The mixture consists of particles with two different size distributions: particles with $r =$\SI{2.12 \pm 0.05}{\micro\meter} (MicroParticles GmbH PS-R 4.2, Batch: PS/Q-R-B1198) and particles with a radii range of $1.0-$\SI{1.5}{\micro\meter} (Spherotech SVP-20-5). The particles are washed prior to use, and their concentrations adjusted to obtain a ratio of about 5:1 (small:large). The larger particles are then selected with the help of the bounding boxes obtained from YOLO and their hydrodynamic radius measured using the Stokes drag method described in the ``Calibration'' section of the Supplementary Materials, moving trapped particles at constant speed in the sample using the motors while measuring the force. To calculate the hydrodynamic radius, we use \cref{eq:radius}.

To perform this experiment, the SmartTrap is instructed to autonomously follow these main steps:
\begin{enumerate}
    \item \textbf{Trap a particle:} Trapping a particle begins by positioning the fluidics chamber using the motors so that the appropriate capillary tube opening is near the optical trap. The microfluidic pump is then used to create a flow of particles from the capillary into the main channel. Once a particle comes into view, the motors move to position the optical trap on the particle. This happens in a loop, so if the particle moves due to the flow from the capillary the target position is updated accordingly. If there are several particles in view, the closest one is chosen. Once a particle is trapped, the flow is turned off and the trapped particle is brought back to the pipette.
    \item \textbf{Select large particles:} If the trapped particle is smaller than a predefined threshold, the system immediately proceeds to step 4 to release it. The size is obtained from the prediction of the YOLO bounding box. This is checked after trapping the particle, since determining the size of particles far from focus is challenging.
    \item \textbf{Measure the hydrodynamic radius:} The sample is moved between predetermined positions using the motors. This drags the trapped particle through the fluid and generates a drag force. From this force, the hydrodynamic radius is calculated using \cref{eq:radius}. The relative fluid velocity is determined from the recorded motor movement speed.
    \item \textbf{Drop the particle:} The trapped particle is released by flushing the central chamber with a strong flow. The system then returns to step 1 to repeat the process and measure the radius of another particle.
\end{enumerate}

\begin{figure}[H]
    \centering
    \includegraphics[width=0.95\linewidth]{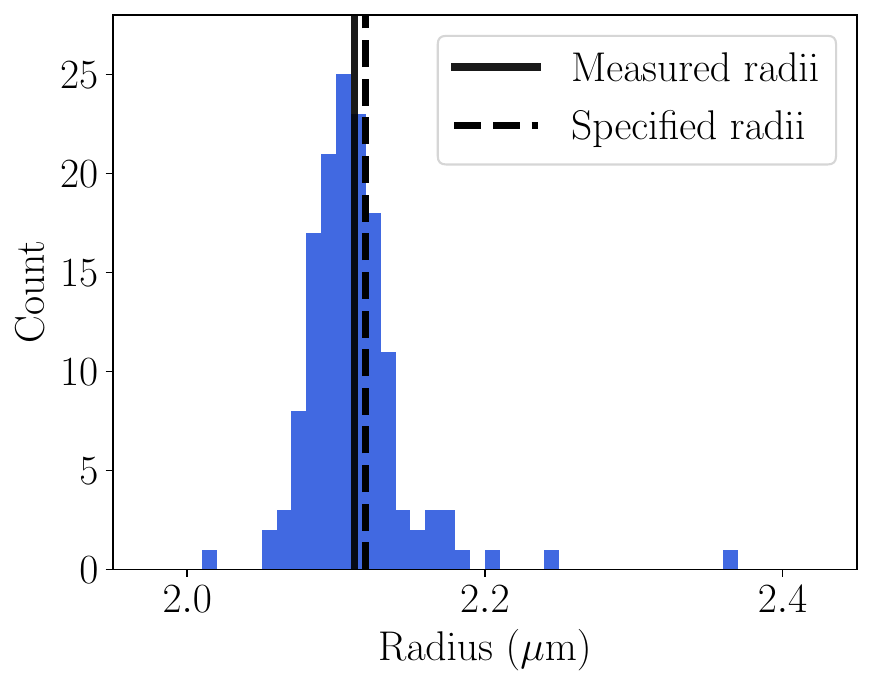}
    \caption{\textbf{Size distribution of particles.} The hydrodynamic radius was measured to be \SI{2.11\pm0.04}{\micro\meter} where the error is the standard deviation of the measurements. The average radius (solid line) and the manufacturer’s specified value (dashed line) are shown in the histogram.}
    \label{fig:SortingResults}
\end{figure}

We test SmartTrap with this protocol for a sample and have it run continuously for 4.5~h. During this time 938 particles were trapped, 159 of which were large. Of these, 15 measurements failed due to trapping more than one particle leaving 144 for the analysis. The results of the analysis are shown in \cref{fig:SortingResults}.
The autonomous protocol was terminated when the system ran out of particles.

There is good agreement between the measured radius of \SI{2.11 \pm 0.04}{\micro\meter} and the manufacturer’s specification of \SI{2.12 \pm 0.05}{\micro\meter}, demonstrating that the selection algorithm effectively separates the larger particles from the smaller ones. The selection is fast, typically taking less than a second to determine if the correct particle has been trapped and dropping it shortly after this. The total throughput depends on the parameters of the system (duration of measurement and concentrations of particles being key parameters); in the measurement reported in \textbf{\cref{fig:SortingResults}}, more than 200 particles were investigated per hour and the hydrodynamic radius was measured for about 30 of these.

\begin{figure*}
    \centering
    \includegraphics[width=\textwidth]{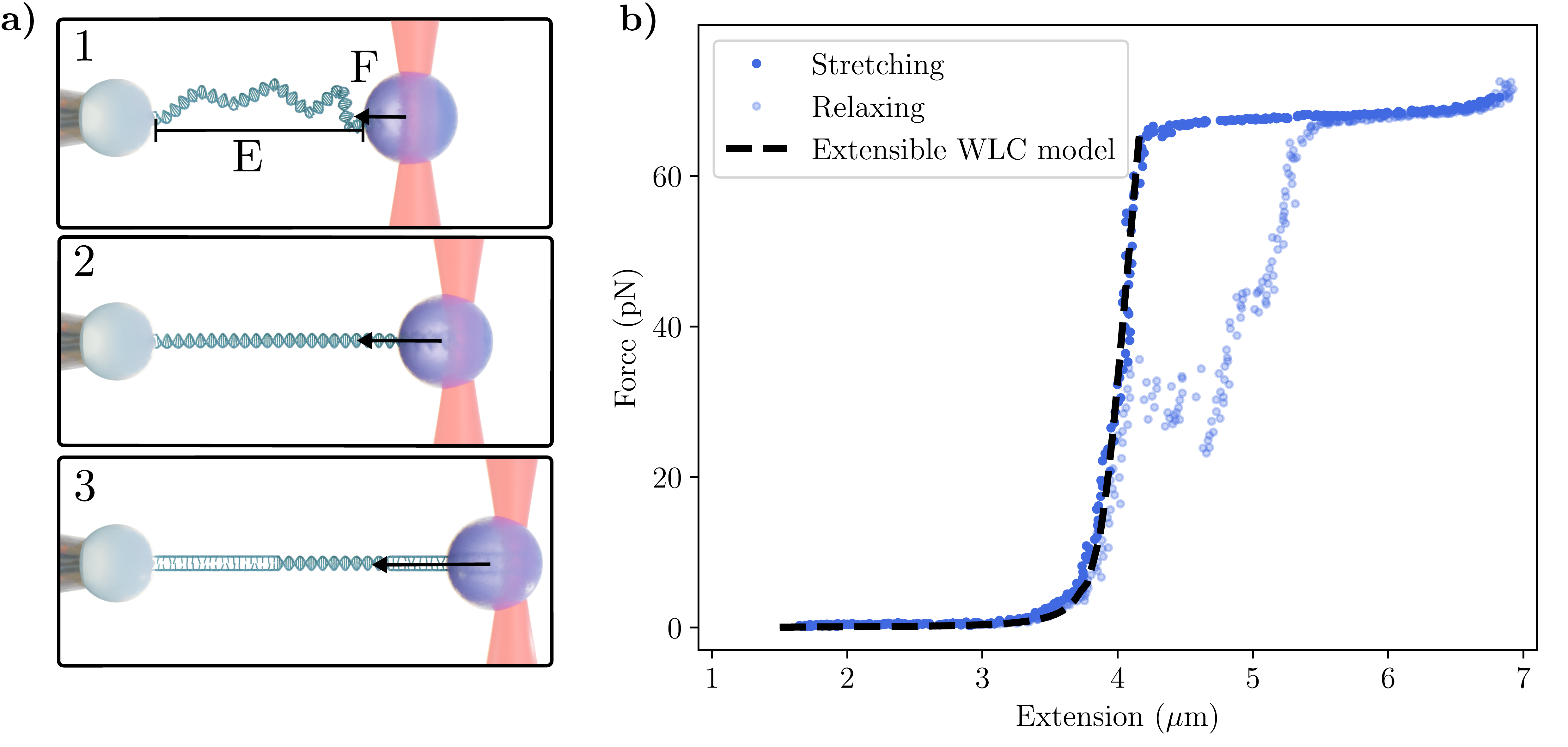}
    \caption{
    {\bf DNA pulling experiment.}
    ({\bf a}) Illustration of a DNA pulling experiment. The DNA molecule is attached between two particles. One of the particles is kept fixed by a micropipette, while the other is moved back and forth using an optical trap. 
    ({\bf b}) Force--extension curves from the experiment (circles) and the extensible WLC model (dashed line). Tracking of the trapped particle is performed during the experiment with the help of the YOLO algorithm. Both the stretching (dark blue circles) and relaxation of the molecule (bright blue circles) are shown. The stretching and relaxation overlap, apart from where there is hysteresis showing as an early drop of the force during the relaxation.
    At small extensions, the molecule exerts a low force and is slightly coiled (panel 1 in {\bf a}).
    As the extension approaches the contour length of the molecule, the molecule straightens out causing the force to increases sharply with the distance (panel 2 in {\bf a}); this behavior is well described by the extensible WLC model.
    Above approximately \SI{65}{\pico\newton}, the molecule overstretches (panel 3 in {\bf a}).}
    \label{fig:DNA-stretching}
\end{figure*}

\subsection*{DNA pulling}

Understanding the structural and mechanical properties of biomolecules is essential to mapping their roles in biological systems. Single-molecule techniques have been transformative in probing transient states—such as protein and nucleic acid folding—and in revealing physical properties (e.g., stiffness and binding interactions) typically obscured in bulk measurements, thereby illuminating processes like DNA replication, transcription, and chromatin organization ~\cite{heller2014optical,neuman2008single,dame2006bacterial,bustamante2003ten}.
Single-molecule force spectroscopy experiments specifically reveal how molecular conformations change in response to an applied force. Optical tweezers are especially well suited for these measurements because of their high spatial and temporal resolution. Among the most fundamental such experiments is DNA pulling, in which the extension dynamics of a  double-stranded DNA molecule is measured as force is applied. This experiment helped establish that DNA can be accurately modeled using the extensible worm-like chain model \cite{smith1996overstretching,bustamante1994entropic}. Although the details vary across different experiments, the core steps—tethering the molecule between two particles, controlling the distance between them, and measuring the force as a function of extension—are shared by many single-molecule experiments, making the DNA pulling assay an archetypical force-spectroscopy measurement.

A DNA pulling experiment is performed by tethering a DNA molecule (in our experiment, a fragment of $\lambda$-DNA molecules with a length of \SI{4}{\micro\meter}) between two particles and measuring the force acting on the DNA molecule while the particles are pulled apart, as schematically illustrated in \textbf{\cref{fig:DNA-stretching}a}.
One particle is held still by a micropipette while the optically trapped particle is moved back and forth.
The optically trapped particle acts also as a force probe. 

The autonomated DNA pulling experiment starts with a configuration routine executed by the user (step 0 in \textbf{\cref{fig:AutomationLoop}c}).
First, the user indicates to the software the positions of the pipette and the side capillaries by manually locating their openings in the chamber. These positions are saved with a single click as motor encoder positions.
The user then needs to indicate which channel of the pump is connected to which channel of the fluidics chamber; this tells the system which pump channel to activate to start the flow of particles of a specific type.

Once the initial configuration is finished, the autonomated DNA pulling experiment consists of six steps, as illustrated in  \textbf{\cref{fig:AutomationLoop}c}:
\begin{enumerate}
    \item \textbf{Check pipette and trap:} The algorithm determines whether a particle is present in the optical trap or in the pipette by checking if any particle is located within a prescribed distance of the estimated laser position or the pipette tip, respectively. Assuming that it is the start of the procedure, there will be nothing in either the trap or the pipette.
    
    \item \textbf{Trap streptadavin particle:} The program traps a streptadavin-coated particle (SA particle), which will be held by the pipette. This is done in the same way as in the particle characterization experiment, but importantly the system goes to the capillary connected to the channel with SA particles.

    \item \textbf{Suck into pipette:} With a SA particle trapped and positioned near the pipette, the next step is to catch it with the pipette. The tip of the pipette is located using the YOLO algorithm. The trapped particle is then approached to the tip, first using the motors for rough alignment, and then the wigglers to finely adjust the laser position. Once the particle is within a few microns from the tip, the pump connected to the pipette is briefly turned on creating a flow into the pipette and sucking the particle firmly into the tip. Thereafter, the program goes back to the saved position near the pipette to ensure that the particle is in the pipette and not in the trap; if it is not, it will try again.

    \item \textbf{Trap DNA particle:} Next, the particle with DNA molecules is trapped using the same method as for the first (SA) particle but instead going to the capillary containing the particles with DNA.
    
    \item \textbf{Attach DNA:} Once a SA particle is in the pipette and a DNA particle is trapped near the pipette, it is time to get the DNA to attach to the SA particle. This is done by first positioning the trapped particle above the one in the pipette using the motors and position feedback. Next, the system aligns the focus of the two particles by adjusting the z-position of the chambers so that the z-positions of the particles, as predicted by the convolutional network, match. Then, they are aligned, also in the x direction (perpendicular to the pipette) with the help of the wigglers which provide superior precision to the motors.

    After this initial alignment, the particle in the trap is gently approached to the one in the pipette using the wigglers. This is done until there is a weak repulsive force between the particles (about \SI{5}{\pico\newton}), indicating that the trapped particle is being pushed out of the trap by the SA particle in the pipette. Thereafter, the particles are separated, also using the wigglers. If things went well, then one DNA molecule on the trapped particle will have attached to the SA particle in the pipette when they were close.

    To check if a molecule is attached, the two particles are separated and if there is a significant force at large particle--particle separation, this indicates that a molecule is attached. For the cut $\lambda$-DNA molecules, this means a force of at least \SI{60}{pN} at a distance of \SI{4}{\micro\meter} or more. This is because we expect a force plateau at approximately \SI{65}{\pico\newton} for extensions greater than \SI{4}{\micro\meter} where DNA overstretches \cite{smith1996overstretching}. These numbers can be adjusted to better suit other molecules, providing an easy pathway to automate other similar experiments.

    \item \textbf{Perform experiment:} Finally, a pulling protocol is loaded onto the micro-controller. The protocol defines how far apart the particles should be moved and at what speed. The micro-controller provides feedback at \SI{7}{\kilo\hertz}, much faster than the communication from the host computer, which is why this protocol is run directly on the micro-controller rather than the host offering smoother motion of the lasers. An upper force threshold is used to find the upper limit of the separation, and a minimum separation distance between the particles is used for the lower limit.

    Once the protocol is started, the program also starts recording data. The protocol will repeatedly separate the particles to stretch the molecule and thereafter let it relax. This is repeated for a user specified duration of time, which we have set to 3 minutes. 
    
    If the molecule, or one of the tethers, is broken before the 3 minutes have elapsed, the protocol and data recording will be stopped and the system will revert to trying to attach a molecule. If the molecule does not break, the protocol will stop automatically when the time limit is reached. The particle is then removed by flushing the central chamber with buffer solution for a few seconds and the experiment is repeated with another particle. Because there are no molecules attached to the SA particle, it does not need to be replaced. However, if the flow is sufficiently strong, the SA particle may also be lost, in which case it is replaced by a new SA particle.
\end{enumerate}
How the various steps work in practice is shown in Supplementary Video 3.

In addition to the general procedure described above, there are also continuous checks to detect if something goes wrong throughout the experiment and autonomous procedures to handle most such situations. For example, not all particles have DNA attached, which is why the program will only try to attach the DNA for a limited number of times (10) before trying with a different particle.
If instead two DNA molecules get attached, this will show up in the subsequent analysis as a sharper increase in the force than what we expect, but it does not otherwise change the procedure and is dealt with in the data analysis after the experiment is finished. 
The presence of more than one particle in the trap is detected by checking if the predicted z-position deviates from what it is with only one particle in the trap. The program also checks that the trapped particle is not lost throughout the experiment. Both of these are rare events which are handled by going back to the starting state and, in the case of double trapped particles, by also dropping what is in the trap and flushing the central channel to get rid of excess particles. 

We have run SmartTrap with the DNA pulling protocol for several hours without supervision during which more than a dozen particles were tested---see Supplementary Video 3. 
The number of experiments performed per hour depends on the specific experiment with the likelihood of molecule attaching, particle concentrations, and pulling rate having a large influence. With the settings we employed, where each experiment is terminated after 3 minutes and the particle replaced, the system performs experiments on 10-15 DNA molecules per hour. This means that it typically takes 3 minutes to release the previous particle, trap a new one, align it, attach a molecule and perform the pulling protocol.

The experiment yields the force as a function of the extension, as shown in \textbf{\cref{fig:DNA-stretching}b}.
The tracking is performed in real time during the experiment with the use of the YOLO network, which limits the need for post-processing of the data. Up until the overstretching, the force--extension curve can be modeled by an extensible worm-like chain model (extensible WLC, \cref{eq:eWLC_model}).
In \textbf{\cref{fig:DNA-stretching}b}, we see that there is an excellent agreement between the model and the data (see the ``DNA Pulling Experiment'' section in the Supplementary Materials). Since video tracking provides the position of the particles rather than molecular extension,  
and since we do not know the precise molecular attachment point on the fixed pipette particle, 
our data have been offset to align with the model and give the same extension for a force of \SI{20}{\pico\newton}. There is also some hysteresis when the particles are moved back due to force-induced melting of the molecule. The hysteresis is seen as early drops in the force curve.
Lastly, there is a force plateau at about \SI{65}{\pico\newton} where the molecule overstretches. The plateau is approximately \SI{2.8}{\micro\meter} long, corresponding to $70\%$ of the contour length of the molecule, which is in good agreement with previously reported results for $\lambda$-DNA \cite{smith1996overstretching}.
The results were also consistent between different molecules as showed in \textbf{Suppl. \cref{fig:DNA_comparison}}. 

\subsection*{Stretching of red blood cells}

The mechanical properties of red blood cell membranes are vital to their biological function, as these cells must deform to navigate through narrow capillaries and withstand the shear forces present arising during blood circulation \cite{omori2012tension}. Changes in membrane stiffness are associated with conditions such as sickle cell disease and malaria \cite{diez2010shape}. Moreover, because red blood cells lack most internal organelles and cytoskeletal structures, relatively small forces can induce significant deformation, making them highly sensitive probes for mechanical studies \cite{guo2014microfluidic,guck2001optical}. Therefore, understanding how red blood cell membranes respond to force not only provides insight into fundamental biophysical processes but also has clinical implications for diagnosing and treating diseases linked to abnormal cell rigidity.

To stretch red blood cells, we exploit the fact that the momentum of light changes when light passes between two media with different refractive indices, such as the interior of a cell and the surrounding buffer. This gives rise to a force acting on the interface that is directed normally to it and away from the denser media. For a trapped cell, the force is away from the cell membrane both when the light enters and exits the cell. By trapping using various laser intensities, we are able to see a clear difference in the shape of the cells depending on the trapping power. The higher the trapping power, the more elongated the cells are along the propagation axis, giving them a smaller cross section when viewed from the camera. Our approach is similar to the one used in the optical stretcher device \cite{guck2001optical}.

\begin{figure*}
    \centering
    \includegraphics[width=0.9\textwidth]{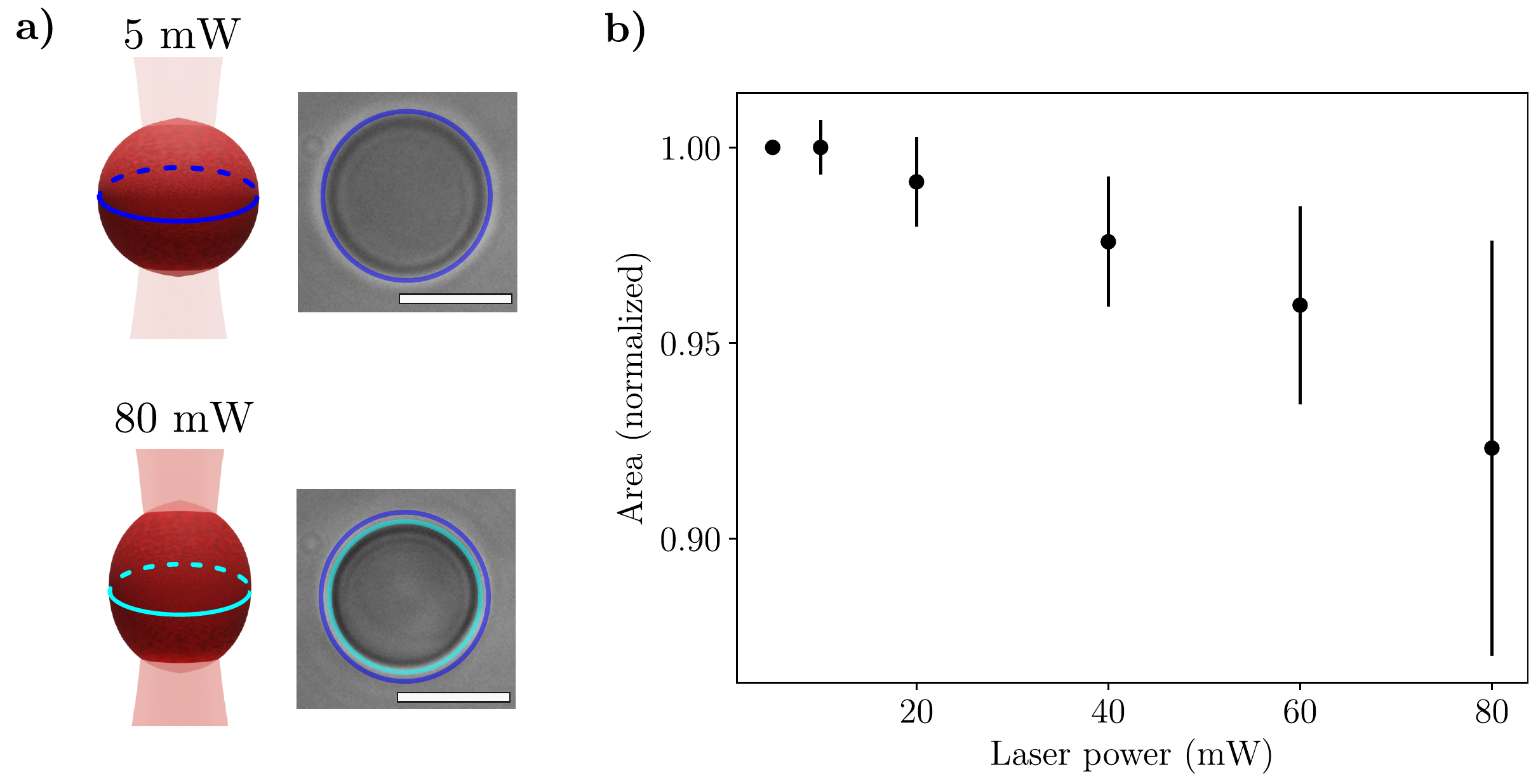}
    \caption{
    {\bf Optical stretching of red blood cells.} 
    ({\bf a}) Illustrations of
    a trapped red blood cell. When the trapping power is \SI{5}{\milli\watt} per laser, there is no significant stretching and the cell is near spherical due to the low osmotic pressure. To the right, there is an experimental image with the dashed blue line showing the outline of the cell. When the cells are trapped with \SI{80}{\milli\watt} per laser, they stretch along the propagation axis of the laser. In the experimental image in the bottom right, there is the outline from the same cell as above when trapped in low power, highlighting that the cell has shrunk in the transversal plane. The scale bar is 5 microns.
    ({\bf b}) The cross sectional area of the cells gradually decreases with increasing trapping power corresponding to their shape becoming more prolate. The error bars show the standard deviation of the relative areas, ten red blood cells from the same sample were used with an average cross-sectional area of \SI{39 \pm 4}{\micro\metre\squared} as measured when trapped at \SI{5}{\milli\watt} power. The units are normalized by the size of the cells when trapped at \SI{5}{\milli\watt}. }
    \label{fig:RBCillustration+results}
\end{figure*}

For these experiments, human red blood cells are diluted in a low osmolarity buffer, making them inflate and become nearly spherical. The solution is then flown into the microfluidic chamber of the SmartTrap where the red blood cells are trapped. Initially, the lasers are set to a low power (approximately \SI{5}{\milli\watt} in the sample from each laser). Then, the power of the two traps is changed simultaneously in steps up to \SI{80}{\milli\watt}. Trapping at low power establishes a baseline at which the cells are almost perfectly spherical. At higher laser powers, this force is greater, which stretches the cells more, decreasing their cross sectional area; this stretching is illustrated in \textbf{\cref{fig:RBCillustration+results}}{\bf a}.

These experiments are comparatively easy to automate since they involve only a single cell and therefore do not make use of the micropipette. The SmartTrap just needs to trap a red blood cell and then record its profile while the trapping power is changed. The procedure can be split into four main steps and is shown in Supplementary Video~4:
\begin{enumerate}
    \item \textbf{Flow cells into the chamber:} The flow in the central channel is briefly turned on, with the aim of bringing cells into view and removing cells from previous measurements from the optical trap.
    \item \textbf{Look for cells:} The SmartTrap looks for cells within the field of view. If there are no visible cells, it reverts back to step 1 to flow more cells.
    \item \textbf{Trap a cell:} With red blood cells in view, the system proceeds to trap the closest one. The laser power is set to \SI{5}{\milli\watt} so as not to deform the cells during trapping.
    \item \textbf{Measure cross section:} The trapping power is changed in steps while the transversal profiles of the cells are recorded in separate videos, one for each power for subsequent analysis. Once the measurement is finished, the system will go back to step 1 and the flow will guarantee that the same cell is not measured twice.
\end{enumerate}

Performing measurements on a single cell takes 2 to 3 minutes. Since the cells are nearly spherical and homogeneous, they are very similar in appearance to the particles used in the DNA experiment, the primary differences being a lower refractive index and greater size. Therefore, by just including a few dozen cells in the training data, the YOLO algorithm is able to detect them. To quantify the stretching, the cross-sectional area of the cells is monitored by the camera and measured using standard imaging techniques, specifically a Gaussian filter followed by a threshold which extracts the area of the cells.

As can be seen in \textbf{\cref{fig:RBCillustration+results}}{\bf b}, the cells contract markedly with increasing laser power. Due to the differences in trapping geometry, a direct comparison between different stretching methods is difficult. Yet, our results and those of \cite{guck2001optical} are of similar magnitude for the same laser power. We also note that the cross-sections of the cells decrease continuously with increasing laser power.

\subsection*{Electrostatic interaction between particles}

In colloidal sciences, it is essential to measure the interaction forces between particles (e.g., electrostatic, Van der Waals, and hydrodynamic interactions) to understand phenomena such as self assembly, adsorption, and aggregation \cite{cosgrove2010colloid,adamczyk1996role}. These interactions have broad industrial applications, for instance, in the food industry, pharmaceutical industry, and water treatment. There are multiple methods for assessing these interactions, many of which look at bulk solutions, such as dynamic light scattering (DLS) and electrophoretic mobility measurements. Optical tweezers opened up the possibility of studying these interactions at the single particle level and also on the exact same particles in different conditions (e.g., temperature, salinity, pH). Electrostatic interactions, in particular, are central to stabilizing colloidal suspension by preventing particles from getting close enough to aggregate \cite{cosgrove2010colloid}.

\begin{figure*}
        \centering
    \includegraphics[width=\textwidth]{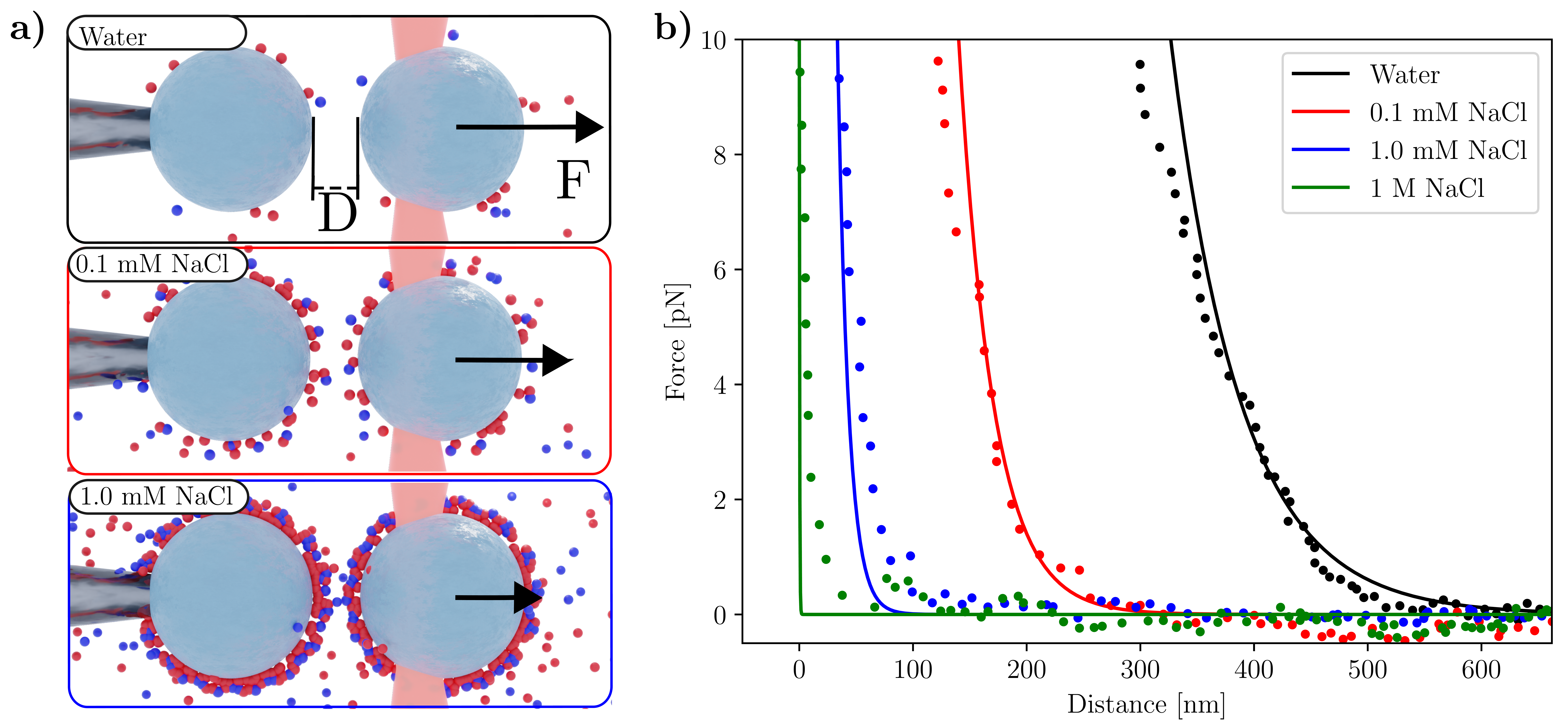}

    \caption{
    {\bf Electrostatic interactions between particles.}
    ({\bf a}) Depiction of the electrostatic interaction experiments across varying salt concentrations. As the salt concentration increases, so does the screening which reduces the electrostatic force, illustrated as more ions aggregated around the particles. 
    ({\bf b}) Force as function of particle-particle distance for various salt concentrations. The experimental data (dots) agrees well with a fit of the electrostatic force as described by DLVO theory (solid lines).}
    \label{fig:ElectrostaticExperiment}
\end{figure*}

To measure the electrostatic repulsion between particles in an optical tweezers, two particles are brought close to each other, and their positions and the force acting on them are measured simultaneously at varying distances. The configuration, shown in \textbf{\cref{fig:ElectrostaticExperiment}}{\bf a}, is similar to that used in \cite{gutsche2007forces}. Because the particles carry a small stabilizing charge from the sulfate end groups on their surfaces, the electrostatic force will repel the particles from one another at short distances. By measuring this repulsive force as a function of distance, we map this interaction, repeating the process across a range of different salt concentrations changes the interaction force.

SmartTrap can autonomously measure the electrostatic repulsion between multiple particle pairs in a single medium. The protocol used for this is similar to that of the DNA pulling, steps 1-3 being the same, but with some key differences in the later steps related to the measurements:
\begin{enumerate}
    \item \textbf{Check the pipette and trap:} see step 1 in the DNA pulling protocol.

    \item \textbf{Trap a particle:} see step 2 in the DNA pulling protocol.
    
    \item \textbf{Suck into pipette:} see step 3 in the DNA pulling protocol.
    
    \item \textbf{Trap the second particle:} A second particle, of the same type as the first is trapped and brought next to the particle already in the pipette.
    
    \item \textbf{Align the particles:} The trapped particle and the one in the pipette are aligned by ensuring that they are in the same focal plane and have the same x-position. The alignment is similar to that performed in the DNA pulling experiment when attaching DNA.
    
    \item \textbf{Measure the repulsion:} The particles are pushed together until the force exceeds a certain value, typically \SI{10}{\pico\newton}. This is used as the first endpoint of the protocol. The second endpoint is set to a little less than a micron away from the first where the electrostatic force is negligible. Also, the lasers are moved at a speed of less than \SI{10}{\nano\meter\per\second} to avoid any hydrodynamic effects from the particles moving close together.
    
    Once finished, the system will flush the central chamber with a strong flow to clear both the pipette and trap from particles.
\end{enumerate}

This protocol in action is shown in Supplementary Video~5. When comparing results across different salt concentrations, it is crucial to use the same particle pair. Variations in particle size and surface charge would otherwise make measurements incomparable, given the short range of electrostatic forces. Manually closing off the channels and replacing the solution without losing the particles ensures the integrity of the experiment. Our autonomous system still plays a vital role in real-time tracking, force measurements, and 3D alignment, while also helping to determine appropriate flow rates for solution replacement. 

We start with distilled water and incrementally increase the salt concentration by flushing the chamber for 30 minutes at each step.
The ions from the salt screen the electrostatic interaction, meaning that for a certain distance the repulsive force will be lower at higher salt concentrations, as illustrated in \textbf{\cref{fig:ElectrostaticExperiment}}{\bf a}. 
In this experiment, the two particles get very close to one another, a situation that is challenging since the images of the particles overlap making standard methods fail \cite{baumgartl2005limits}. We found that the YOLO algorithm yields too noisy results in this case. Because of this, the particles are tracked using a U-Net which has been trained on simulated data, \textbf{Suppl. \cref{fig:UnetData}}, and specifically to handle the case of particles being very close, see the ``U-Net Model'' section in the Supplementary Materials.

From \textbf{\cref{fig:ElectrostaticExperiment}}{\bf b}, we see a good agreement between experiments and theory. It also becomes apparent why it is essential to use the same pair of particles to see the differences; the difference between high and low salinity is largely that the interaction appears at a longer distance and for salt concentrations above ca \SI{0.1}{\milli\mol} this distance is on the order of \SI{100}{\nano\meter}, less than one twentieth of the particle diameter. The increase in the force is not as sharp as theory predicts for the two highest salt concentrations. We attribute this to the sulfate end groups on the particle surfaces contributing with a slight repulsive force.

\section*{Discussion}

Measuring forces and mechanical properties with optical tweezers is still largely a manual process, limiting the amount of data that can be collected. SmartTrap represents a significant step toward overcoming this limitation. The experiments presented here cover a wide range of common optical tweezers applications. Particle characterization, although relatively simple in our demonstration, has a wide range of potential applications (e.g., selecting certain types of cells on which to perform experiments) that could be readily implemented by modifying the selection criteria from size to another feature. DNA pulling serves as a paradigmatic single-molecule force spectroscopy experiment. With some parameter tuning, it could be adapted to measuring other molecules (e.g., proteins, DNA hairpins). It would then be ideal for mapping the energetics of molecular processes across different conditions. One could, for instance, measure DNA nearest-neighbor energies for various temperatures or salinities. This type of high-accuracy experiments is central to understanding molecules and require a large number of measurements \cite{Rissone2025,huguet2010single}. In contrast, the red blood cell experiment proved easier to automate than DNA pulling, yet clearly illustrates that one can, with modest effort, efficiently perform a large number of simple measurements on individual cells or particles. The electrostatic repulsion experiments, on the other hand, exemplify the measurement of interparticle forces. Notably, the algorithm used there is essentially a simplified version of that used for DNA pulling, illustrating how the steps building up a complex protocol can be repurposed for other applications. This makes subsequent automation protocols quicker and easier to implement.

Compared to a human operator, SmartTrap generally matches or exceeds manual operation in both measurements per hour and overall experiment quality. We attribute this advantage to faster feedback and more precise measurements of relative positions. As the particle characterization experiment demonstrates, when measurements can be performed quickly, the system is capable of conducting a large number of measurements autonomously in a short period. The fact that the autonomous protocols can reliably run for an extended time greatly reduces the time needed by researchers to perform experiments. Furthermore, having the instrument working on its own also reduces the risk of human bias from inconsistent operation, potentially very important when trying to observe rare events. Still, there are scenarios in which human operators still outperform the system, such as detecting particles that are significantly out of focus or moving very rapidly. These situations benefit from detecting motion by comparing consecutive frames. YOLO operates on a frame-by-frame basis and thus struggle with fast movement and identifying particles moving far from focus, particles which are detectable to humans. It is likely that a more specialized tracking algorithm could outperform humans also in these scenarios.

Many of the challenges faced during automation stem from having one of the particles held in the pipette, which introduces an additional point of failure and adds complexity by requiring precise 3D alignment. This requirement largely explains why preparing a new measurement is significantly faster in the particle characterization experiment than in the DNA pulling experiment, even when the particle in the pipette is not replaced. An alternative approach would be to use two separate traps to ensure both particles remain in the same plane, effectively removing the need for extensive alignment. However, this comes at the cost of increased instrument complexity and would likely require trapping two particles for each subsequent measurement. Despite these constraints, errors during autonomous operation are rare, as demonstrated by the system’s ability to perform both extended measurements and large-scale studies on numerous particles, molecules, or cells. This reliability is primarily due to solutions already in place for common problems, such as trapping two particles, losing the particles, or failing to attach DNA. When issues do occur, they usually stem from external experimental factors that require human intervention such as running out of particles or the pipette clogging. These issues are relatively simple to fix but currently require manual intervention.

Looking further ahead, the trend of handing over repetitive work to computers is likely to continue. As the breadth of the autonomous protocols we have implemented demonstrates, large parts of protocols can often be reused in other protocols, greatly lowering the barriers to automation. Although our system is an optical tweezers, several building blocks of the autonomous functionality could be used to perform other smart microscopy experiments. For instance, the trapping of particles, seemingly highly specific to optical tweezers, works by actively centering particles in the field of view. Meaning that, without modification, it could be used to follow the path of single particles and by retraining the YOLO network, as we did for the red blood cells, it could instead follow following single plankton or bacteria in a sample. As the technique matures a wider range of experiments will be automated. Nonetheless, it is unlikely that humans will be entirely removed from the process anytime soon. Especially in the early stages when starting a new experiment and designing protocols, manual operation is still necessary before handing control over to the computer. However, we have found that some of the functions of the autonomous system, such as automatic alignment of particles and real-time tracking in 3D, are very useful also when manually testing or performing new protocols. It is therefore likely that such hybrid modes of operation will become the norm, assisting researchers in both established and novel experimental endeavors.

\newpage
\bibliography{bibliography}

\clearpage

\appendix
\section*{Supplementary Material}
\renewcommand{\thefigure}{S\arabic{figure}}
\setcounter{figure}{0}

\section{Figures}

\begin{figure}[H]
    \centering
    \includegraphics[width=\linewidth]{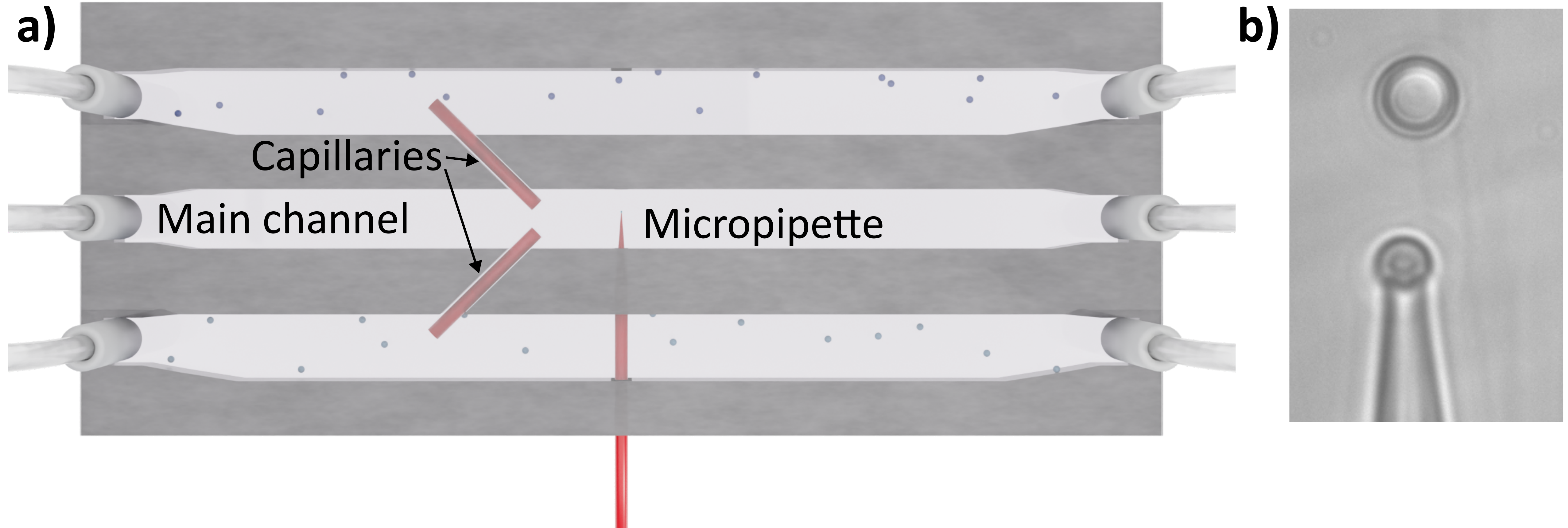}
    \caption{\textbf{Microfluidics chamber.}
    (\textbf{a}) Illustration of the microfluidics chamber. There are three separate channels. The middle channel is used to perform the experiment and contains the tip of the micropipette where one of the particles is held. The bottom and top channels are used to flow particles into the main channel via glass capillaries. The micropipette and the capillaries are colored in red for clarity. The channels are connected to a microfluidics pump with independent operation.
    (\textbf{b}) Picture obtained from an experiment while it is being set up: there is one particle in the optical trap (top) and one in the pipette (bottom).}
    \label{fig:MicrofluidicsChamber}
\end{figure}

\begin{figure}[H]
    \centering
    \includegraphics[width=\linewidth]{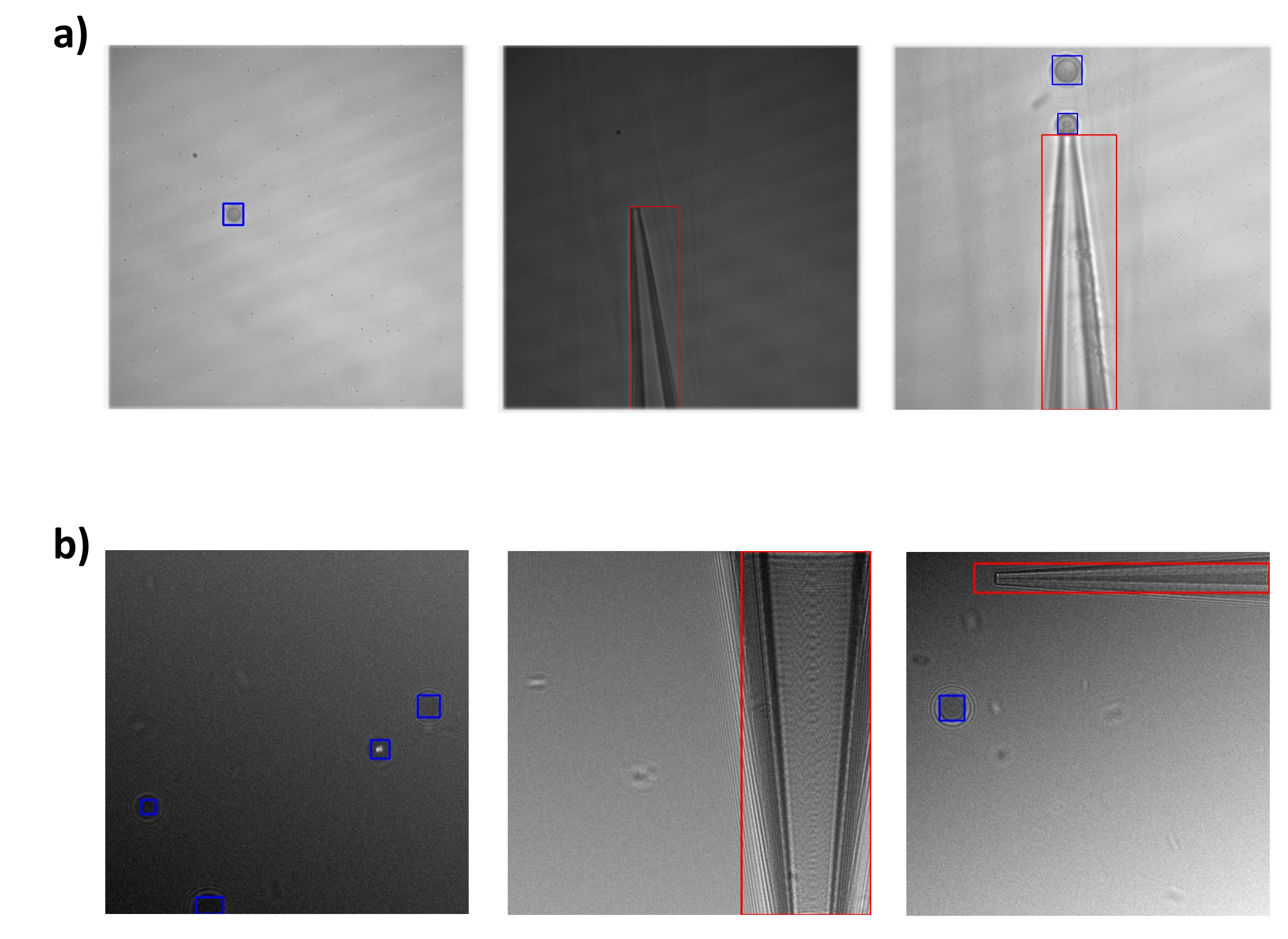}
    \caption{\textbf{YOLO training data.} Example of training data used for the YOLO network. (\textbf{a}) Manually annoted data from experiments and (\textbf{b}) simulated training data obtained with the help of the DeepTrack2 Python package. The blue boxes highlight the particles and the red box highlights the pipettes.}
    \label{fig:YOLO_training}
\end{figure}

\begin{figure}[H]
    \centering
    \includegraphics[width=0.95\linewidth]{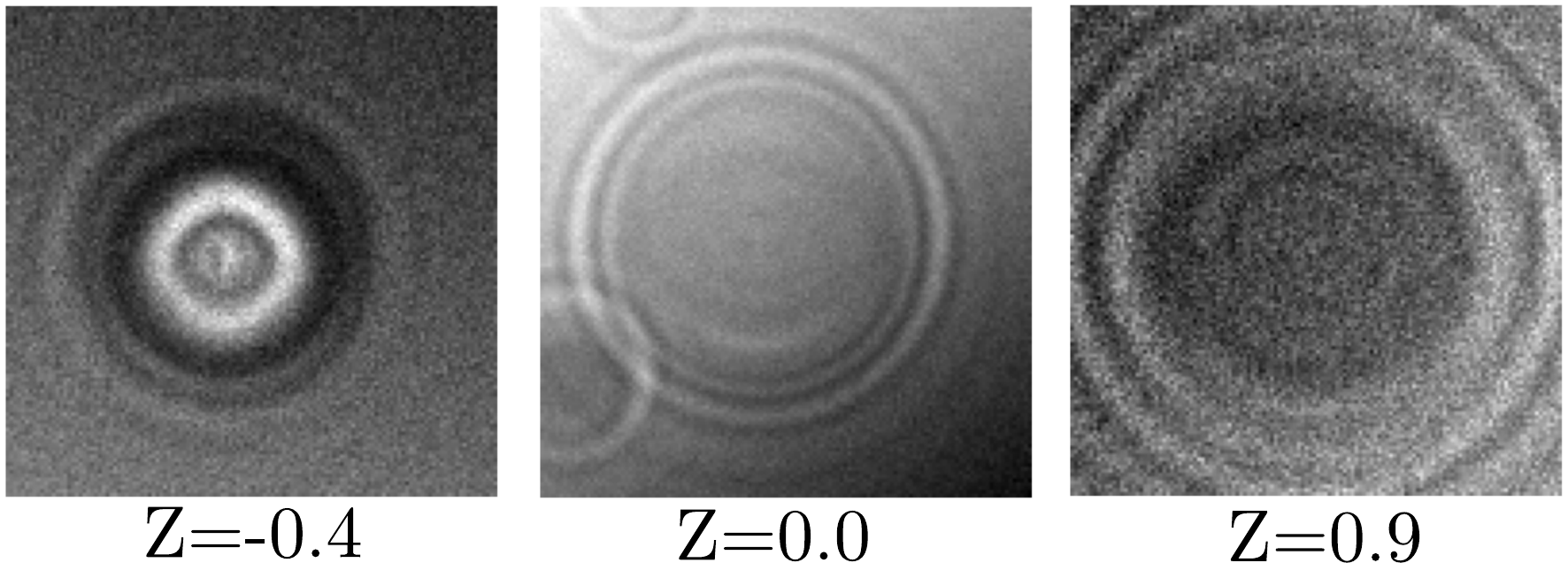}
    \caption{\textbf{Simulated particles with different focus.}
    Example of simulated images used for training the convolutional neural network to predict the focal position of particles. The numbers below the pictures indicate the target z-value.}
    \label{fig:Z_trainingData}
\end{figure}

\begin{figure}[H]
    \centering
    \includegraphics[width=0.95\linewidth]{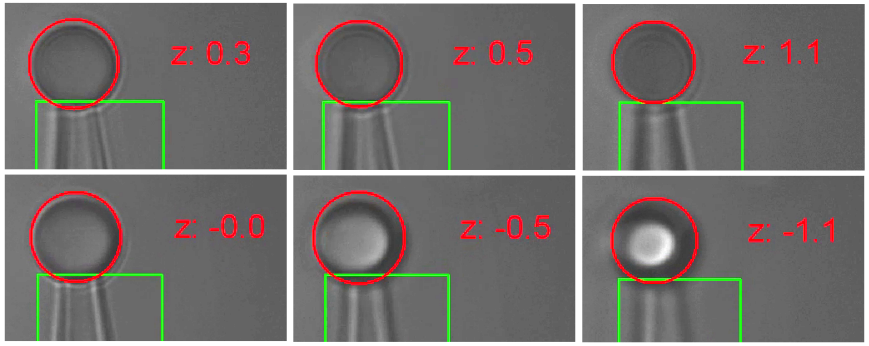}    \caption{\textbf{Experimental predictions of the particle and pipette positions.}
    Screenshots from the program predicting the lateral positions of the particle (red circle) and of the pipette (green box). 
    It also predicts the axial position (z-position) of the particle relative to the focus, written in red.}
    \label{fig:Z-focuse_real}
\end{figure}

\begin{figure}
    \centering
    \includegraphics[width=0.95\linewidth]{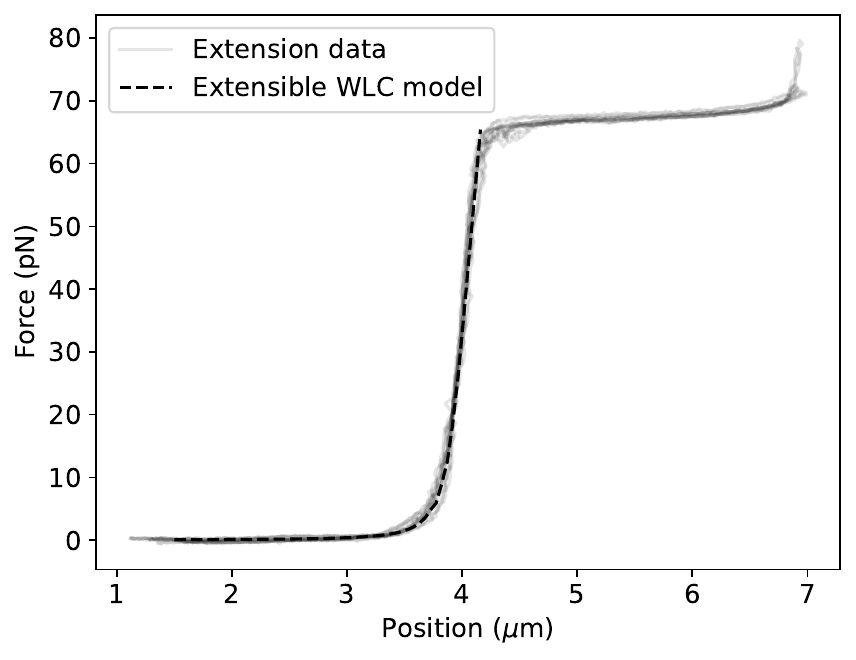}
    \caption{\textbf{Multiple different force--extension curves from autonomous experiments.}
    14 different curves from 7 different molecules aligned to have the same position when the force is \SI{20}{\pico\newton}. These curves were all obtained from the same experiment session and include all experiments in which a single DNA attached properly (just one DNA attaching and tethers not breaking before reaching the overstretching plateau).}
    \label{fig:DNA_comparison}
\end{figure}

\begin{figure}[h]
    \centering
    \includegraphics[width=\linewidth]{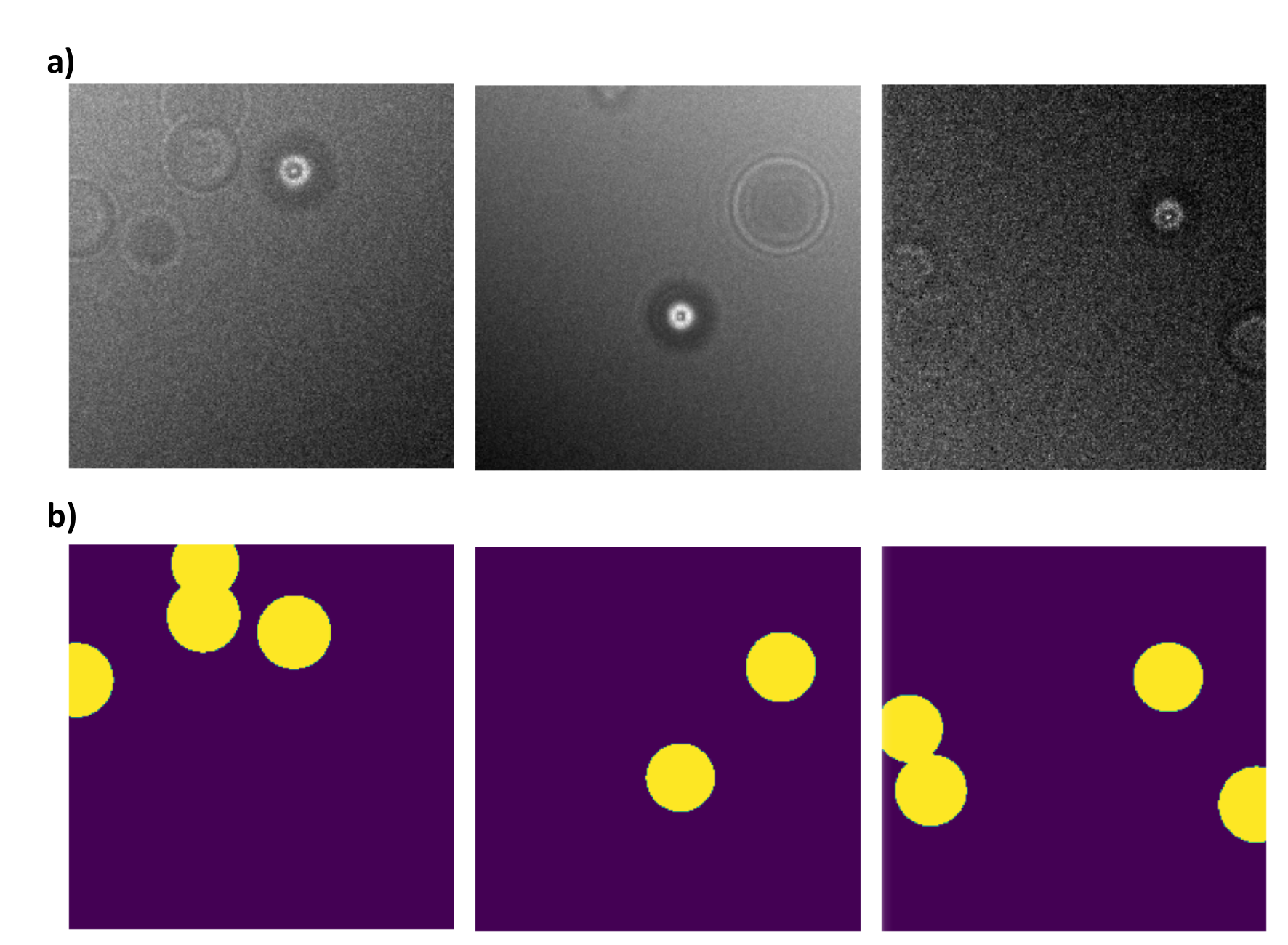}
    \caption{\textbf{U-Net training data.}
    (\textbf{a}) Simulated images generated using  DeepTrack2 and used for training the U-Net. Small particles are added to the background and used as noise to prevent false positives in the output (e.g., classifying the pipette as a particle).
    (\textbf{b})The targets are binary masks with the same size as the images. The U-Net is tasked to predict 1 for pixels containing particles and 0 otherwise.}
    \label{fig:UnetData}
\end{figure}

\newpage

\section*{Optical tweezers system}

The optical tweezers system used is a counter-propagating tweezers inspired by the MiniTweezers design originally developed by the group of Carlos Bustamante at UC Berkeley~\cite{tweezerslab}. 
Importantly, we have designed our own custom circuit boards, firmware, and graphical user interface to get full digital control of the instrument.
Having digital control of all the components of the system in a single software proved essential for automation. The code and electronic schematics for the system are available from our GitHub repository ~\cite{MiniTweezersSoftware}.

\subsection*{Optical paths}

The two lasers (both Lumics lasers, model LU0808M250) follow equivalent but mirrored paths to form the counter-propagating optical trap, as shown in  \textbf{\cref{fig:miniTweezersSchematics}b}. 

Here, we describe the path of laser A.
Laser A exits from an optical fiber into wiggler A, which controls the axial position using a piezoelectric actuator. The function of the wiggler is described in detail in the ``Laser wigglers'' section.
Next, a small portion (about $8\%$ of the laser power) is deflected by a pellicle beamsplitter and focused onto position detector A. This is a PSD used to measure the laser position. The remainder of the laser light is collimated by a lens and passes through a quarter waveplate before entering the back-aperture of the objective. The quarter waveplate ensures that the light is circularly polarized when entering the objective.
The objective focuses the laser inside the sample, creating the optical trap. After passing through the sample, the laser is collected and collimated by the second objective before passing a second quarter waveplate, which again linearly polarizes the light. Both objectives are Olympus UPlanSAPO 60X Water Immersion, NA = 1.2. Importantly, the laser is now polarized so that it is transmitted through the first polarizing beamsplitter and reflected off the second. Finally, laser A is focused by a lens onto force detector A (also a PSD). This PSD is positioned in a pivot point relative to the center of the sample ensuring that the laser can be moved in the sample without giving a reading on the force sensors.

There are also an iris and a photodiode that are used to measure the intensity at the center of the beam. When a particle is trapped in a neutral position, the ratio between the light collected by the photodiode and that collected by the PSD is the same as when there is nothing in the trap. If a trapped particle moves along the z-direction (left and right in the schematics), the particle will act as a lens. Depending on the direction of the displacement, it will either focus or defocus the laser on the photodiode through the iris. This allows for the measurement of momentum changes in the axial direction, which after a calibration can be used to calculate the force along the z-axis \cite{bustamante2006light}.

Laser B follows an equivalent but mirrored path as laser A.

The sample is imaged using a brightfield configuration with a blue LED and a short-pass filter positioned before the camera to filter out any stray light from the lasers.

\subsection*{Force measurements}

The primary reason for using counter-propagating beams is that it allows for direct force measurements~\cite{smith20037}. This is achieved by monitoring the momentum change of the lasers as they pass through the sample. Any change in momentum of the lasers as they pass through the sample will be caused by objects in the trap. By Newton's third law, this momentum change gives rise to a force on the trapped object. Importantly, all the scattered laser light needs to be collected to get an accurate reading of the momentum change and thereby the force. This is achieved by not overfilling the back-apertures of the objectives (i.e., only a small portion of the back-aperture is covered by the laser). Because of this, large-angle scattering does not occur, which means that the objectives are able to collect practically all the laser light passing through the sample. Therefore, the deflection of the lasers is directly proportional to the force acting on the particle.

As an additional benefit, using counter-propagating traps allows for lower NA on the objectives and longer working distances making it easier to work in bulk and thereby avoiding surface effects during measurements.

\subsection*{Laser wigglers}

To move the lasers in the system, a piezoelectric mechanical system called a wiggler is employed. The system is described in US-patent US 7274451 B2~\cite{Wiggler_patent}. The system works by having piezoelectric crystals push on a metal ball that in turn moves the outer metal tube relative to the stationary inner tube, thus gently bending the optical fiber. This provides a simple yet efficient method for moving both the lasers in the sample, allowing for accurate positioning of the lasers and fast feedback algorithms for synchronous movement of the two lasers in the sample.

\subsection*{Microfluidics chamber}

Many experiments require two particles in order to study interactions. To get two particles, a custom microfluidics chamber with a micropipette is used. The chamber has three channels: one central channel for performing the experiments and two side channels through which the functionalized particles are flowed. These side channels are connected to the central channel by glass capillaries through which particles can flow. These glass capillaries have an outer diameter of \SI{100}{\micro\meter} with an inner diameter of \SI{25}{\micro\meter} and are cut to the appropriate length using a scalpel. In the illustration shown in  \textbf{Suppl. \cref{fig:MicrofluidicsChamber}a}, the particles flow from left to right when the pumps are turned on.

The chamber is handmade and consists of two sheets of parafilm and two glass slides. When making a chamber the two parafilm sheets are first cut into the appropriate size using a laser cutter. Simultaneously, the laser cutter cuts out the channels, giving the parafilm the shape seen in \textbf{Suppl. \cref{fig:MicrofluidicsChamber}a}. Holes are also cut with the laser cutter in one of the two glass slides to make inlets and outlets. Next, one of the sheets of parafilm is placed on the glass slide with holes. The holes are aligned with the channels and the capillaries and the micropipette are positioned on the parafilm. The second sheet of parafilm is added on top followed by the second glass slide before the chamber is sealed by warming the parafilm close to its melting point. The micropipette is made from glass capillaries with an outer diameter \SI{80}{\micro\meter} and an inner diameter of \SI{40}{\micro\meter} using a custom capillary puller. The puller uses a platinum filament to heat the glass and a small weight to apply a consistent force to the capillary. The heating of the filament is tuned to give a pipette opening with a diameter of ca \SI{1}{\micro\meter}.

To control the flow in the 3 channels a microfluidics pump system is used. The system is an OB1 from Elvesys with 3 independent pressure controlled pumps. Each is connected to one of the 3 channels of the chamber to give dynamic control of the flows. Lastly a separate pump is connected to the micropipette, this pump is a one-way air pump (D2028B, SparkFun Electronics), which provides motorized suction.

\subsection*{Electronics}

The electronics has been designed to take advantage of the widespread availability of powerful microcontrollers. We use an Arduino Portenta as microcontroller unit (MCU) because it is easy to obtain and program, requiring minimal prior experience and no custom equipment. This makes it both straightforward to connect and program, requiring just a USB-C cable. The controller samples the various photodetectors, steers the laser wigglers, and controls the motors that move the sample.
The schematics of the electronics are, like the software, available from our GitHub page \cite{MiniTweezersSoftware}.

\subsubsection*{Sampling of position sensitive detectors}

All four PSD detectors are handled using the same type of circuit and they are all reverse-biased with \SI{15}{\volt}. The signals are amplified in two stages. The first stage acts as a transimpedance amplifier, converting each of the four current signals from the detectors into voltages. The four signals correspond to $X_1, X_2, Y_1, Y_2$ (two for each axis). In the second stage, the differences $X=X_1-X_2$ and $Y=Y_1-Y_2$ as well as the sum of the y-axis signals $S=Y_1+Y_2$ are obtained using standard amplifier subtraction and addition circuits. These 3 signals are then sampled independently and these give the power and position. The laser power is proportional to the sum signal $S$. The difference signals are proportional to both the displacement and the power of the laser and so is the force, meaning that the $X$ and $Y$ difference signals are directly proportional to the force along the corresponding axis.
To get the laser position, independent of the laser power used, the positions $X_p$ and $Y_p$ signals are extracted as $X_p=\frac{X_1-X_2}{S}$ and $Y_p=\frac{Y_1-Y_2}{S}$. It is possible to perform this division operation using analog circuitry but in order to limit the complexity of the circuit design it is done digitally in our system. 
The photodiodes used for measuring the z-force are also reverse-biased. Similarly to the PSDs, they generate a current signal which we convert to a voltage signal using a standard transimpedance amplifier circuit and sending the signal directly to the AD converter.

All the signals are sampled using a 16-bit 16-channel analog-to-digital converter (ADC, model AD7616BSTZ-RL from ANALOG DEVICES). 

For controlling the piezoelectric actuators that steer the lasers, a high voltage amplifier is used that yields an output in the range $0-$\SI{150}{\volt}, making \SI{75}{\volt} the center position. The amplifiers are controlled by a digital-to-analog converter (DAC). The DAC is a 4-channel 16-bits model which is interfaced using the SPI protocol from the microcontroller allowing for fast and accurate control.

A motor driver circuit L293D provides the motors with the driving signal and their direction is controlled with two digital pins. The Arduino Portenta uses software interrupts that trigger on the movement of the motors which updates an internal counter, increasing it if the motor moves forwards, and otherwise decreasing it. A Proportional–integral–derivative (PID) feedback algorithm is used to get stable motor speed when moving.

\subsection*{Software}

A custom software suit has been designed and implemented to control the instrument. It can be broadly split into four main parts: the first is the firmware, which is written in C and runs on the microcontroller; the second is the communication software; the third is the Graphical User Interface (GUI); and the fourth is the set of automation algorithms customized for each type of experiment (described in the results section of the main text). The communications software, GUI and automation algorithms are all written in Python and run on the host computer.

\subsubsection*{Firmware}

The firmware runs on the microcontroller and communicates with the instrument hardware. It reads from the ADC, writes to the DAC, and controls the motors. The sampling of the PSDs is triggered every \SI{64}{\micro\second} by a software interrupt procedure resulting in a sample rate of \SI{15.625}{\kilo\hertz}. A USB-serial protocol is used to continuously send data to the host computer through the built-in USB-C port of the Arduino Portenta.

\subsubsection*{Communications software}

The communications software is what enables the host computer to read data from the instrument and send commands to it. Since this needs to happen continuously, it is important that the Global Interpreter Lock (GIL) of Python does not interrupt the process. Similarly, it is important that the communications does not interrupt any other processes, such as capturing images using the camera. Therefore the communication is handled by two separate classes both of which run asynchronously. The first class is run in a separate process (core on the computer) and sends data to and from the instrument using the serial USB protocol. It always sends the most recent command to the instrument and the data being read is continuously put into a Python queue. The second class runs in a separate thread and processes data from the queue in chunks. First, the data are unpacked by converting from bits to integers and sorting the data depending on which signal it corresponds to. Thereafter forces and positions are calculated, see Supplementary section Calibration.

\subsubsection*{Graphical User Interface}

The GUI is designed to control and monitor the communications with the instrument controller in a user-friendly manner. Enabling features such as clicking and dragging in the live view of the sample to position the lasers and to move the motors. We use a Basler ace a2A5320-23um camera but any camera from this manufacturer will work without any modifications to the code, as will cameras from Thorlabs (other cameras can be integrated with minor modifications to the code).
The GUI also works as a simple oscilloscope, enabling the user to plot and compare signals from the various sensors in real-time. This is most readily used to monitor the forces on the particle during an experiment. The plotting tool also allows for real-time data processing such as running an FFT algorithm or averaging the data.
The GUI also has features such as real-time tracking in the live-feed as well as the ability to draw forces acting on the trapped particle. Examples of the live-plotting and drawing of forces are shown in Supplementary Video 3 where also a screen recording of the experiments is included.
This allows users to identify where the DNA has attached on the particle in the pipette.

\subsubsection*{Automation algorithms}

The automation algorithms are run in the background of the GUI in a separate thread taking care of steps 2,3 and 4 of the automation loop in \textbf{\cref{fig:AutomationLoop}a}. Running in a separate thread prevents other operations (e.g. data image capturing) from having to wait for the automation algorithms when for instance a particle detection is performed. The user can decide which autonomous algorithm to run from the interface turning them on/off dynamically. Also, single procedures can be toggled individually, such as trapping a particle or alignment of the pipette, to help assist users when manually operating the instrument. The details of the various decision procedures are outlined in the main text.

\subsection*{Automatic alignment of counter-propagating traps}

In order for a counterpropagating optical trap to be stable, the two lasers need to be perfectly aligned in the sample. This can be very challenging if the optical trap is also to be moved since any small difference in the optical path will cause the lasers to drift apart in the axial position when moved. Therefore, a feedback algorithm is employed to keep the two lasers in the same axial position. This algorithm makes use of the fact that if the two lasers are aligned, and a particle is trapped, then they will have the same force reading.
In practice the force PSDs are set to zero in software when there is no particle in the trap. This compensates for the fact that the lasers may not be hitting the exact center of the sensors. 
Then, when a particle is trapped, if the beams are not perfectly aligned in the x-y plane, the two beams will be deflected by the particle equally but in opposite directions. This is compensated for by the feedback which moves the lasers to ensure that the PSDs give the same force reading as described in \cite{smith20037}. This feedback is run at ca 7 kHz meaning that the lasers follow each other very well also when moved, as during the DNA pulling protocol.

\subsection*{Calibration}\label{sec:calibration}

The calibration of the optical trap is essential for performing measurements with the tweezers, because it enables the conversion of voltage readings in bits to physical forces and distances. These conversion factors are given by the proportionality constants for the force PSDs and for the position PSDs. How these proportionality constants are obtained is outlined below. However, before these constants are obtained, a fundamental calibration is performed using a micrometer ruler mounted instead of the microfluidics chamber. This gives a calibration factor relating pixels on the camera to microns in the sample.

To calibrate the force PSDs, we exploit the Stokes' law which relates the viscous drag on a particle to the flow velocity at low Reynolds number \cite{stokes1851effect}. Since the experiments are carried out in a microfluidics chamber, we need to account for the distance to the two walls, which is why we use the corrected formula from \cite{smith1992direct}:
\begin{equation}\label{eq:stokes}
    \vec{F_{\rm d}}=-6\pi\mu r \vec{v}(1+2\frac{9r}{16d})
\end{equation}
In this formula, $\vec{F_{\rm d}}$ is the drag force, $\mu$ is the dynamic viscosity, and $\vec{v}$ is the velocity of the particle. The experiments are carried out in the center of the chamber, which has two walls separated by ca \SI{200}{\micro\meter}, which gives a value of $d=$\SI{100}{\micro\meter}. In our case, the viscous drag is generated by moving a trapped particle back and forth using the stage motors. 
The drag force displaces the particle in the optical trap which in turn gives readings on the PSD detectors, where $P_{\rm r}$ is directly proportional to the force, $\vec{F_{\rm d}}\propto P_{\rm r}$. The reading is directly proportional to the force because the change in laser light momentum is the same as the force exerted on the trapped particle, and the dual-lateral PSD measures transverse light momentum directly.  Unlike single-beam traps using backfocal-plane interferometers which cannot collect all the deflected light, our force calibration factor $P_{\rm r}$ remains constant regardless of changes in particle size, shape, or refractive index \cite{smith20037}.

With the instrument calibrated, the Stokes drag method can be used to measure the size of particles. Then, \cref{eq:stokes} is solved for $r$ which gives:
\begin{equation}\label{eq:radius}
    r = \frac{-1}{2b} + \sqrt{\frac{|\vec{F_{\rm d}}|}{ab}+\frac{1}{4b^2}}, a = 6\pi\mu |\vec{v}|, b = \frac{18}{16d}
\end{equation}
In the case of the particle characterization experiments, we use the tabulated value for $\mu$ for water at room temperature (\SI{21}{\celsius}) of \SI{0.9775}{\milli\pascal} \cite{NIST_webbook}.

We exploit the digital camera to calibrate the position readings of the lasers. This is done autonomously with a  trapped particle, moving it in a square pattern using the wigglers while recording its position with the real-time tracking. Then, the particle position is fitted to the signals, correcting for any differences in alignment or sensitivity of the detectors. It is important to note that the position signal is the laser position, rather than the particle position, and these may differ significantly when there is a force acting on the particle. Because of this, the position, as obtained from the camera, is sometimes preferred over the PSD readings, even though the sampling rate of the PSD signal is much greater than that of the camera. 

\section*{Artificial neural networks}
Artificial neural networks are computational models that form the foundation of most modern artificial intelligence algorithms. They get their name from their loose resemblance of biological neural networks \cite{DLCC,Goodfellow-et-al-2016}. There are multiple different artificial neural network models, referred to as architectures, which are adapted to various tasks. Here we give a short outline of the networks used in the SmartTrap system to enable automation and highly accurate particle tracking.

\subsection*{YOLO to predict lateral position of particles and pipette}

A YOLO (You Only Look Once) network was employed for object detection, specifically YOLO v5s, which is an updated and smaller, and therefore faster, version of the original YOLO architecture. This network is available as a software package in Python \cite{yolov5}.

YOLO networks are trained to predict rectangles enclosing objects in an image, known as bounding boxes. YOLO networks can be trained to differentiate between a wide range of different object classes. In our case, we are only interested in two classes, namely, ``pipette'' and ``particle''. 

During training, the network aims to minimize a loss function that contains three components: localization, classification, and objectness. The localization loss measures the accuracy of the location of the bounding box (center and size) quantified with mean squared error. The classification loss is used to teach the network which class of objects are present in the image, and it uses binary cross entropy to evaluate how accurately the class is predicted. Lastly, the objectness (or confidence loss) assesses how confident the network is in its predictions. If a box is supposed to contain an object, the loss penalizes low confidence scores (objectness), and conversely, if the box does not contain any object, it penalizes high confidence scores.
For each training image, the network predicts multiple bounding boxes and assigns confidence scores for object presence and class probabilities.

The network was trained using the stochastic gradient descent algorithm on a combination of simulated and manually annotated images (\textbf{Suppl. \cref{fig:YOLO_training}}). In the manually annotated images, the location and size of particles and pipette were marked by hand using images from previous experiments performed with the same system so as to be as similar as possible to the autonomous experiments. They were of varying size and illumination to cover a wide range of experimental conditions. However, because the exact position was not known and annotation was done manually, it risked introducing noise in the dataset, which is why we also included simulated images in the training set. Since our dataset of manually annotated images is comparatively small (ca 1,000 images), we train for only 100 epochs to avoid overfitting.

The network consistently predicts the relative focal position of the particles. The GUI allows for demonstrating this by printing the prediction directly on the screen as shown in Supplementary Video~1 in which a particle in the pipette is moved in and out of focus.

\subsection*{Convolutional network to predict focus position}

To estimate the focal position of the particles, we use Pytorch \cite{PYTORCH} to train a convolutional network that predicts the z-positions of the particles in the images.
By using simulated data we can circumvent the challenges associated with creating an experimental dataset, something which would require accurately measuring the focal position of a very large number of particles \cite{midtvedt2021quantitative,deeptrack2}. 
The simulated images are made to be similar to the experimental images, and by including noise they are made more challenging as shown by the examples in \textbf{Suppl. \cref{fig:Z_trainingData}}. The network is a convolutional neural network with two sets of convolutional layers, each followed by a max pooling layer. A dense layer on top of this handles the final prediction and training uses the mean square error as loss function. Even though all the training data is simulated, the network predicts also the position of real particles accurately, as illustrated in \textbf{Suppl. \cref{fig:Z-focuse_real}} and Supplementary Video~1.

\subsection*{U-Net model for accurate particle tracking}

For accurate tracking of particles that are very close to one another, which is the case in the electrostatic experiments, we use a U-Net \cite{ronneberger2015u}. U-Nets are widely used in biomedical image analysis for segmentation tasks. 
In a typical segmentation setting, the network outputs a binary image that is 1 wherever an object of interest appears and 0 otherwise. For multiple classes, the output becomes a stack of images, one for each class. To obtain the positions of individual particles from the U-Net predictions, we apply a threshold to the output and then compute the centers of the connected regions. This post-processing step yields the coordinates of the particles of interest.

The U-Net architecture gets its name from its characteristic “U” shape: an initial series of downsampling layers progressively reduces the spatial resolution of the input while capturing large-scale features, followed by a corresponding set of upsampling layers that restore the original resolution. Residual (skip) connections bridge matching downsampling and upsampling layers, allowing the network to preserve fine details. Our network has five sets of convolutional downsampling layers with 64, 128, 256, 512 and 1024 filters in the respective layers and the upsampling path is the same but in reversed order (with 1024, 512, 256, 128 and 64 filters).

The training data consists solely of simulated images \cite{midtvedt2021quantitative,deeptrack2} since in this case we need to know the true particle position for accuracy. In the simulated images, the number of particles is large so that they often overlap, see \textbf{Suppl. \cref{fig:UnetData}}. Furthermore, the size of the particles is tuned to resemble that of the experiment. Very small particles are used in the background of the images to simulate structured noise. This was found to eliminate the risk of the network treating the pipette as a particle.

Despite its higher accuracy, there are two reasons not to use the U-Net in the real-time automation. First, it is significantly more computationally demanding than YOLO, which would increase the processing time. Reducing the size of the U-Net would help, but this would also reduce its accuracy. Second, to achieve reliable segmentation of the pipette, we would need to include manually annotated experimental data, meaning that the accuracy would probably not significantly exceed that of YOLO, except when particles are close to overlapping or when there are many particles in view, a situation in which YOLO may struggle but not found in our experiments.

\section*{Experimental details}

All experiments were performed in the same system using the same type of microfluidics. However, each different experiment requires specific preparation of the sample and analysis of data which is outlined below.

\subsection*{DNA pulling experiment}

We used a fragment of $\lambda$-phage DNA obtained by cutting the full 48.5 kb genome with the restriction enzyme EagI to obtain shorter fragments that can be more easily fully extended in our instrument. Cutting the DNA with EagI results is 3 fragments of lengths $19.9$ kb, $16.7$ kb and $11.8$ kb respectively. The experiments presented are performed on the $11.8$ kb fragment, which has a contour length of \SI{4.0}{\micro\meter}. The two ends of this fragment are labeled with digoxigenin and biotin respectively, which enables them to bind specifically to anti-digoxigenin- and streptavidin-coated particles. The anti-digoxigenin particles are ca \SI{3.4}{\micro\meter} in diameter and the streptavidin ca \SI{2.2}{\micro\meter} making it possible to visually distinguish the particles. Importantly, only one strand at each DNA end is anchored, resulting in a lower overstretching force than would occur if both strands were immobilized \cite{van2009unraveling}. The complete DNA preparation protocol can be found in \cite{DNA_prep_protocol}.
The experiments were performed in a high salt buffer (1 M NaCl, 1 mM EDTA, 10 mM Tris-HCl, components purchased from Sigma-Aldrich).
Before the experiment the DNA is attached to the anti-digoxigenin coated particles. This is done by incubating concentrated DNA molecules and particles together for 20 minutes.
This allows for one of the digoxigenin ends to bind to the anti-digoxigenin, leaving the DNA attached to the particles with one free biotin labeled end. During the experiment the biotin end will attach to the streptadavin coated particle held in the micropipette.

The elastic behavior of DNA can be modeled using the worm-like chain model (WLC model) and in particular the extensible WLC model works well until the molecule starts to overstretch \cite{wang1997stretching}. The extensible WLC model can be approximated as
\begin{equation}
F(x) = \frac{k_{\rm b}T}{L_{\rm p}}\frac{1}{4(1-x/L_0+F/K_0)^2}-\frac{1}{4} + \frac{x}{L_0}-\frac{F}{K_0}
\label{eq:eWLC_model}
\end{equation}
where $k_{\rm b}$ is the Boltzmann constant, $x$ is the extension of the molecule, $L_{\rm p}$ is the persistence length, $T$ is the temperature, $K_0$ is the stretch modulus and $L_0$ is the length of the molecule. For the curve in \textbf{\cref{fig:DNA-stretching}}, we use a persistence length of \SI{43}{\nano\meter} and a stretch modulus of \SI{1200}{\pico\newton}, typical values for double stranded DNA \cite{bustamante2021optical}.

\subsection*{Red blood cells preparation}

Human red blood cells are prepared before each experiments by finger pricking. Approximately \SI{5}{\micro\liter} of blood is extracted and diluted in 2 ml of buffer solution consisting of PBS-10XA, diluted by a factor of 25, and 10 mM glucose. This creates a low osmotic pressure environment, which makes the cells inflate, becoming almost spherical.
During the experiments, the two objectives are placed ca \SI{3}{\micro\meter} closer than during normal trapping. There are two main reasons for this small displacement of the objectives: having the two foci of the counter-propagating lasers in different positions reduces heating, while also slightly increasing the stretching. 

\subsection*{Electrostatic repulsion measurements}

The two particles used in the electrostatic experiments are of the same type. They are polystyrene particles with a mean diameter of \SI{4.24 \pm 0.1}{\micro\meter} (MicroParticles GmbH  PS-R 4.2, Batch: PS/Q-R-B1198). 

Electrostatic interactions between colloidal particles can be described by the Derjaguin-Landau-Verwey-Overbeek (DLVO) theory \cite{derjaguin1941theory, verwey1947theory}. DLVO theory provides a framework for understanding the interactions at play between the particles, including the van-der Waals attraction, the electrostatic repulsion and the screening of the repulsion by dissolved ions. For distances longer than a few nanometers, the interaction is dominated by the electrostatic repulsion. The electrostatic force between two charged particles can be described by
\begin{equation}
F(D)= \frac{(e Z)^2}{4 \pi \varepsilon_0 \varepsilon_r } \left( \frac{(1+\kappa(D+2R))\exp(-\kappa D)}{(1+\kappa R)^2 (D+2R)^2 } \right)
\end{equation}
where $e$ is the elementary charge, $Z$ is the number of charge groups per particle, $\kappa$ is the Debye length, $D$ is the surface-to-surface distance between the particles and $R$ is the particles radius \cite{gutsche2007forces}. $\varepsilon_0$ and $\varepsilon_r$ are the permeability of free space and the relative permeability, respectively.
For sufficiently high salinity, the electrostatic repulsion becomes small enough that it can be easily overcome by thermal fluctuations, which enables the particles to come into contact. We exploit this to find the particle radii by doing a final measurement in which the salinity is so high that the particles are brought into contact. This is how we determined a radii of \SI{2.11}{\micro\meter} in average for our particle pair, very close to the value of \SI{2.12}{\micro\meter} specified by the manufacturer. From the fit shown in \textbf{\cref{fig:ElectrostaticExperiment}}, the number of charge groups is found to be $Z \approx 460,000$. We performed the fit only on the 0.1~mM data because it has a smaller relative error in the distance measurement than the 1~mM dataset, and a smaller relative error in $\kappa$ (due to uncertainty in the salt concentration of our milli-Q water) than the measurement in water without added salt.

\section{Supplementary videos}
The supplementary videos are recordings of experiments, either screen recordings or recordings of the camera feed performed with the user interface. The first video shows the tracking and the subsequent videos illustrate the autonomous experiments. The videos are available from our GitHub repository ~\cite{MiniTweezersSoftware}.

\subsection{Supplementary video 1 - 3D tracking}
This video is a screen recording of the camera view in the user interface with real-time tracking visualization turned on. The video shows where the program detects the pipette and the particle. This is done while simultaneously moving the pipette with the particle in and out of focus, estimating the particle focal position.

\subsection{Supplementary video 2 - Autonomous particle characterization}
The video shows the autonomous particle characterization being performed. The video starts with the instrument moving to a capillary to trap a particle. The first particle trapped is one of the target particles, which is determined by the autonomous algorithm based on it being larger than a certain threshold. Since it is one of the target particles the hydrodynamic radius is measured by moving it between two fixed positions in the sample while recording the forces and motor movement. Once the measurement is finished the microfluidics pump connected to the central chamber is turned on creating a flow which removes the trapped particle. Next, another particle is trapped. By chance this is a small particle, which should not be characterized, therefore it is immediately removed, again using the pump connected to the central chamber. Next, a situation in which multiple particles are trapped is shown. In this case the software detects that the focal position is offset compared to what is expected from a single particle and therefore the particles are removed even though the profile size match those of the target particles. 

Lastly, several experiments are displayed at a high speed with a timer added to illustrate how the system can run for extended times characterizing dozens of particles.

The recording is performed using the interface itself rather than recording the screen itself. This gives slightly higher video quality than a screen recording.

\subsection{Supplementary video 3 - Autonomous DNA pulling}
Video illustrating the autonomous DNA pulling. First, a recording of a single autonomous pulling is showed starting with an empty chamber, without particles in either trap or pipette. This part of the video is commented and is a recorded camera feed showing all the different steps of the autonomous pulling process. Starting with checking the pipette, followed by trapping of the streptadavin particle and positioning of this in the pipette. This is followed by the trapping of a particle with DNA which is then attached to the particle in the pipette where after the experiment measurement is perfromed. After this first pulling, the video shows the full graphical user interface while the program performs a large number of pullings autonomously. This part of the video is showed at increased speed (to limit the duration of the video) and also the real-time plotting is shown.

\subsection{Supplementary video 4 - Autonomous red blood cells measurements}
The videos shows how the red blood cell experiments are performed by the instrument. The video is a screen recording of the GUI and also includes a timer. After a cell has been trapped, initially at low power, cells profile is recorded in a video. Then the trapping power is briefly increased and the cell profile at higher power is recorded in another video. Thereafter the trapping power is reduced and the process repeated four more times at increasingly high powers. Thereafter the cell is removed by a flow. Since there are cells dispersed in the medium the flow is also used to bring new cells into view which are trapped and measured. If by chance there are no cells in view after flowing new medium then the flow is briefly turned on again.

The video speed is increased to show more experiments in a shorter time.
\subsection{Supplementary video 5 - Autonomous electrostatic repulsion measurements}
Showcase of how electrostatic repulsions can be measured autonomously. When the video starts there is nothing in either trap or pipette. The program first focuses the pipette and checks its content. Next it moves to the capillary, turns on the microfluidics pump and traps a particle. The trapped particle is then placed in the pipette. After confirming that the particle has successfully transferred to the pipette, a second particle is trapped and brought to the pipette. The particle in the trap is aligned to the particle in the pipette. The trapped particle is pushed towards the particle in the pipette by moving the trap to find appropriate limits for the measurement protocol. Next the protocol and recording of data (both force and video) are started. At this stage the program zooms in on the two particles to limit the size of the videos. Once the measurement is completed, a strong flow removes both particles resetting the experiment and preparing the system for another measurement.

\section{Acknowledgments}
We acknowledge support from the the Horizon Europe ERC Consolidator Grant MAPEI (grant number 101001267), and the Knut and Alice Wallenberg Foundation (grant number 
2019.0079). We would also like to acknowledge the support from the Waernska professorship, ERC Proof of Concept 2024 (Grant agreement ID: 101212879), Adlerbertska stipendier 2023 (grant number AD2023-3065) and Donationsnämndens stipendier 2023 (grant number DS2023-0867).

\section{Author Contributions}
The instrument was constructed and aligned by M.S with the help of A.C and G.P with S.B.S providing guidance. S.B.S developed the original optical tweezers design together with C.B. M.S was responsible for developing all software and developed the electronics with assistance from L.B and G.P and support from S.B.S. 

M.S conceived the experiments together with G.V, A.C, G.P, S.B.S and C.B. M.S set up the instrument to have it perform the experiments and A.C, J.C.S, V.S.R F.R and I.P  provided practical advice and assistance on how to prepare samples and design experiments. S.B.S prepared the DNA. M.S wrote the paper together with A.C and G.V.
All authors reviewed and approved the final paper.
\section{Competing interests}
The authors declare no competing interests.

\end{document}